\documentclass[12pt]{iopart}

\usepackage{iopams}

\expandafter\let\csname equation*\endcsname\relax

\expandafter\let\csname endequation*\endcsname\relax
\usepackage{amsmath,amsfonts,amssymb,mathabx}
\usepackage{stackrel}
\usepackage[dvipsnames]{xcolor}
\usepackage[pdftex]{graphicx}
\usepackage[pdftex,colorlinks=true]{hyperref}
\hypersetup{
	linkcolor=NavyBlue,
	citecolor=ForestGreen,
	urlcolor=Maroon	
}

\usepackage{placeins}
\usepackage{subfig}
\usepackage{caption}
\newcommand{\R}{\mathbb{R}}

\newcommand{\Z}{\mathbb{Z}}

\newcommand{\map}{\mathfrak{m}}

\begin{document}

\title[Precision measurements of Hausdorff dimensions in 2d quantum gravity]{Precision measurements of Hausdorff dimensions in two-dimensional quantum gravity}

\author{Jerome Barkley and Timothy Budd$^1$}

\address{$^1$ Radboud University, Nijmegen, The Netherlands.}
\ead{t.budd@science.ru.nl}

\begin{abstract}
Two-dimensional quantum gravity, defined either via scaling limits of random discrete surfaces or via Liouville quantum gravity, is known to possess a geometry that is genuinely fractal with a Hausdorff dimension equal to $4$. 
Coupling gravity to a statistical system at criticality changes the fractal properties of the geometry in a way that depends on the central charge of the critical system.
Establishing the dependence of the Hausdorff dimension on this central charge $c$ has been an important open problem in physics and mathematics in the past decades.
All simulation data produced thus far has supported a formula put forward by Watabiki in the nineties.
However, recent rigorous bounds on the Hausdorff dimension in Liouville quantum gravity show that Watabiki's formula cannot be correct when $c$ approaches $-\infty$.
Based on simulations of discrete surfaces encoded by random planar maps and a numerical implementation of Liouville quantum gravity, we obtain new finite-size scaling estimates of the Hausdorff dimension that are in clear contradiction with Watabiki's formula for all simulated values of $c\in (-\infty,0)$.
Instead, the most reliable data in the range $c\in [-12.5, 0)$ is in very good agreement with an alternative formula that was recently suggested by Ding and Gwynne.
The estimates for $c\in(-\infty,-12.5)$ display a negative deviation from the latter formula, but the scaling is seen to be less accurate in this regime.
\end{abstract}

\begin{figure}[h]
	\makebox[\textwidth][c]{
	\includegraphics[width=1.12\linewidth]{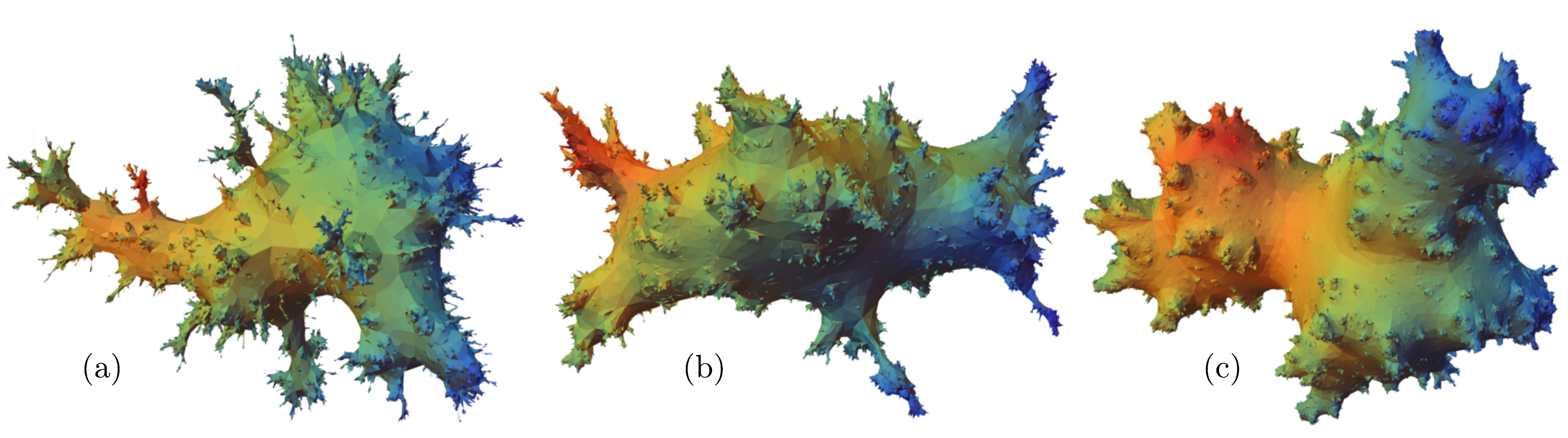}
	}
	\caption{Three-dimensional visualizations of random surfaces sampled from three of the models under consideration: (a) spanning-tree-decorated quadrangulations, (b) bipolar-oriented triangulations, (c) Schnyder-wood-decorated triangulations. Each one consists of $2^{15}$ faces and is coloured (from red to blue) by increasing graph distance to a randomly sampled vertex.
	\label{fig:simulations3d}}%
\end{figure}
\clearpage

\section{Introduction}

The emergence of scale-invariance at criticality and the universality of the corresponding critical exponents is at the center of statistical physics, underlying for instance the self-similar features of spin clusters in the Ising model at critical temperature.
In a purely gravitational theory, where the configuration of a system is encoded in the spacetime geometry, criticality and self-similarity may well be realized in a ultramicroscopic limit if the theory possesses a non-trivial ultraviolet fixed point.
The presence of self-similarity generally means a departure from smooth (pseudo-)Riemannian geometry and forces one to reconsider the local structure of spacetime geometry.
In particular, the dimension of spacetime is no longer a clear-cut notion, but depends on the practical definitions one employs and the fractal properties of the geometry to which these definitions are sensitive. 
Although fractal dimensions have been studied in many quantum gravity approaches (see e.g.\ \cite{Ambjorn2005,Benedetti2009,Reuter2011,Calcagni2015,Carlip2017}), putting such computations on a rigorous footing is difficult.
Arguably the main reason for this is the lack of explicit examples of scale-invariant statistical or quantum-mechanical ensembles of geometries, a prerequisite for the existence of exact critical exponents and fractal dimensions.
Two-dimensional Euclidean quantum gravity forms an important exception in that it provides a family of well-defined such ensembles as well as a variety of rigorous mathematical tools to study them.
This makes it an ideal benchmark in the computational study of fractal dimensions in quantum gravity, albeit a toy model.

Generally speaking, two-dimensional quantum gravity, the problem of making sense of the path integral over metrics on a surface, can be approached from two directions, roughly categorized as the continuum Liouville quantum gravity approach and the lattice discretization approach.
In the first case, one tries to makes sense of the two-dimensional metric as a Weyl-transformation $\rmd s^2 = e^{\gamma \phi} \rmd\hat{s}^2$ of a fixed background metric $\rmd\hat{s}^2$, where the random field $\phi$ is governed by the Liouville conformal field theory with coupling $\gamma\in (0,2]$. 
In the latter case one considers random triangulations (or more general random planar maps) of increasing size in the search of scale-invariant continuum limits.
The famous KPZ relation \cite{Knizhnik1988} describing the gravitational dressing of conformal matter fields was derived in Liouville quantum gravity but shown to hold for lattice discretizations in many instances, providing a long list of exact critical exponents.
These exponents, however, only indirectly witness the fractal properties of the metric. 

That fractal dimensions can differ considerably from the topological dimension was first demonstrated by Ambj\o rn and Watabiki \cite{Ambjorn1995a}, where the volume of a geodesic ball of radius $r$ in a random triangulation was computed to scale as $r^4$ (later proven rigorously \cite{Chassaing2004,Angel2003}), suggesting a Hausdorff dimension of $4$ for the universality class of two-dimensional quantum gravity in the absence of matter. 
The self-similar random metric space of this \emph{Brownian} universality class was later established \cite{LeGall2013,Miermont2013} in full generality as the continuum limit of random triangulations, and was recently shown \cite{Miller2016} to agree with Liouville quantum gravity at $\gamma=\sqrt{8/3}$.
While many geometric properties of the Brownian universality class are known, much less can be said about the universality classes with $\gamma\neq\sqrt{8/3}$ which are supposed to describe scale-invariant random geometries in the presence of critical matter systems.
Notably, the \emph{spectral dimension}, a fractal dimension related to the diffusion of a Brownian particle, has been argued \cite{Ambjorn1998a} to equal $2$ for the full range of $\gamma\in(0,2)$, which has recently been proven rigorously \cite{Rhodes2014,Gwynne2017}.
The dependence of the Hausdorff dimension on the coupling $\gamma$, however, has been a wide open question since the nineties.

\begin{figure}[t]
	\centering
	\includegraphics[width=\linewidth]{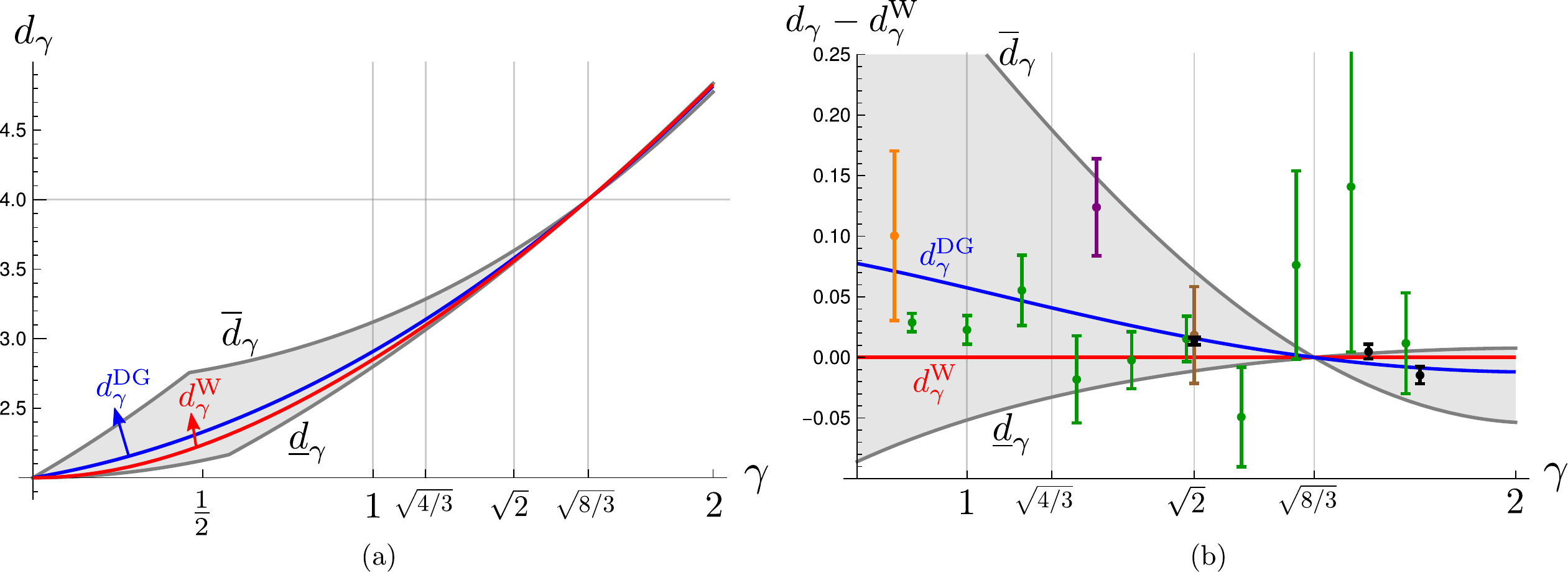}
	\caption{(a) The known bounds \cite{Gwynne2019,Ding2018,Ang2019} on the Hausdorff dimension $d_{\gamma}$ together with the predictions by Watabiki and Ding \& Gwynne. (b) A zoomed-in version shifted by Watabiki's prediction together with numerical estimates: \cite{Ambjorn1995} in brown, \cite{Anagnostopoulos1999} in purple, \cite{Ambjorn2012a} in orange, \cite{Ambjorn2013} in black, \cite{Ambjorn2014} in green.\label{fig:bounds}}
\end{figure}

A formula for the Hausdorff dimension was put forward by Watabiki \cite{Watabiki1993} based on a heuristic computation of heat kernels in Liouville quantum gravity. 
Written in terms of $\gamma$, which is related to the central charge $c$ via \eqref{eq:cgamma}, it reads
\begin{equation}\label{eq:watabiki}
d_{\gamma}^{\mathrm{W}} = 1+ \frac{\gamma^2}{4}+\sqrt{\left(1+\frac{\gamma^2}{4}\right)^2+\gamma^2}
\end{equation}
and correctly predicts $d_{\sqrt{8/3}}^{\mathrm{W}}=4$ for the Brownian value and $d_{\gamma}^{\mathrm{W}}\to 2$ in the semi-classical limit $\gamma\to 0$.
However, its main support came soon after from numerical simulations of spanning-tree-decorated triangulations, a model of random triangulations in the universality class corresponding to $\gamma=\sqrt{2}$.
For this model the Hausdorff dimension was estimated at $d_{\sqrt{2}} = 3.58 \pm 0.04$ \cite{Ambjorn1995} (and $d_{\sqrt{2}} \approx 3.55$ in \cite{Kawamoto1992}), in good agreement with the prediction $d_{\sqrt{2}}^{\mathrm{W}} = \tfrac12(3+\sqrt{17})\approx 3.56$.
Since then all numerical estimates reported for these and other models are consistent with Watabiki's prediction \cite{Ambjorn1995,Anagnostopoulos1999,Kawamoto2002,Ambjorn2012a,Ambjorn2013}.
The most accurate estimates are summarized in Table~\ref{tbl:previousresults}.
These values, as well as measurements from Liouville quantum gravity \cite{Ambjorn2014}, are plotted in Figure \ref{fig:bounds}b.

\begin{table}[h]
	\caption{\label{tbl:previousresults}Previous estimates of the Hausdorff dimension.}
\begin{indented}
	\item[]
	\begin{tabular}{cccrp{1mm}ll}\hline
		Reference & $c$ & $\gamma$ && $d_\gamma$ & & $d_\gamma^{\mathrm{W}}$ \\\hline 
		\cite{Ambjorn2012a} & $-20$ & $0.867\ldots$ & $2.76\!\!\!$ & $\pm$ &$0.07$&$2.6596\ldots$\\
		\cite{Anagnostopoulos1999} & $-5$ & $1.236\ldots$ & $3.36\!\!\!$ & $\pm$ &$0.04$&$3.2360\ldots$\\
		\cite{Ambjorn2013} & $-2$ & $\sqrt{2}$ & $3.575\!\!\!$ & $\pm$ &$0.003$&$3.5615\ldots$\\
		\cite{Ambjorn2013} & $1/2$ & $\sqrt{3}$ & $4.217\!\!\!$ & $\pm$ &$0.006$&$4.2122\ldots$\\
		\cite{Ambjorn2013} & $4/5$ & $\sqrt{10/3}$ & $4.406\!\!\!$ & $\pm$ &$0.007$&$4.4207\ldots$\\\hline
	\end{tabular}
\end{indented}
\end{table}

However, recent mathematical developments in Liouville quantum gravity have shown that \eqref{eq:watabiki} cannot be correct for small values of $\gamma$.
Ding and Goswami \cite{Ding2018a} have proven a lower bound on the Hausdorff dimension $d_{\gamma}$ of the form 
\begin{equation}\label{eq:boundsmallgamma}
d_\gamma \geq 2 + C \frac{\gamma^{4/3}}{\log \gamma^{-1}}
\end{equation}
for some unknown constant $C>0$ and sufficiently small $\gamma$, which is seen to be incompatible with $d_{\gamma}^{\mathrm{W}} = 2 + O(\gamma^2)$.
How is it possible that an incorrect formula agrees so well with numerical data (in some cases at the three digit accuracy)?
There are several conceivable explanations: 
\begin{enumerate}
	\item The fractal dimension we call the Hausdorff dimension in the case of triangulations measures something different compared to the one in Liouville quantum gravity.
	\item The numerical estimate of the Hausdorff dimension in random triangulations is inaccurate.
	\item The actual Hausdorff dimension is not given by \eqref{eq:watabiki}  but close enough in the tested regime to be compatible with the data.
\end{enumerate}
The first option has been ruled out in a very precise sense in the works \cite{Ding2018,Dubedat2019,Gwynne2019a} (and references therein).
In particular it is shown that Liouville Quantum Gravity possesses a unique realization as a scale-invariant random metric space for each value $\gamma\in(0,2)$.
The Hausdorff dimension $d_\gamma$ of this space is a strictly increasing, continuous function of $\gamma$.
Moreover, the Hausdorff dimension of various models of random triangulations, including spanning-tree-decorated triangulations and all other models considered in this paper, are shown to agree with this $d_\gamma$ (with the value of $\gamma$ depending on the universality class).
This warrants us speaking about \emph{the} Hausdorff dimension $d_\gamma$ without specifying the precise model.

Although only the value $d_{\sqrt{8/3}} = 4$ is known, rigorous bounds have recently been derived \cite{Gwynne2019,Ding2018,Ang2019}: $\underline{d}_\gamma \leq d_\gamma \leq \overline{d}_\gamma$ with
\begin{align}
\underline{d}_\gamma &= \begin{cases}
2+\frac{\gamma^2}{2} & 0 < \gamma \lesssim 0.576 \\
{\displaystyle\frac{12-\sqrt{6}\gamma+3\sqrt{10}\gamma+3\gamma^2}{4+\sqrt{15}}} & 0.576 \lesssim \gamma \leq \sqrt{8/3} \\
\frac{1}{3}\left(4+\gamma^2+\sqrt{16+2\gamma^2+\gamma^4}\right) & \gamma \geq \sqrt{8/3},\end{cases}\label{eq:dlower}\\
\overline{d}_\gamma &= \begin{cases}
2+\frac{\gamma^2}{2} + \sqrt{2}\gamma & 0 < \gamma \lesssim 0.460 \\
\frac{1}{3}\left(4+\gamma^2+\sqrt{16+2\gamma^2+\gamma^4}\right) & 0.460 \lesssim \gamma \leq \sqrt{8/3} \\
{\displaystyle\frac{12-\sqrt{6}\gamma+3\sqrt{10}\gamma+3\gamma^2}{4+\sqrt{15}}} & \gamma \geq \sqrt{8/3}.\end{cases}\label{eq:dupper}
\end{align}
Contrary to \eqref{eq:boundsmallgamma}, these bounds are still consistent with Watabiki's formula, see Figure \ref{fig:bounds}a.
Note that for $\gamma=\sqrt{2}$ the bounds imply $d_{\sqrt{2}} \in (3.550,3.633) \subset d_{\sqrt{2}}^{\mathrm{W}} + (-0.012,0.072)$, so deviations from $d_{\sqrt{2}}^{\mathrm{W}}$ will only show up in the third digit.
This lends credence to explanation (iii), but does make one wonder whether it is a coincidence that a relatively simple formula like \eqref{eq:watabiki} precisely fits the bounds \eqref{eq:dlower} and \eqref{eq:dupper}.
However, in \cite{Ding2018} Ding and Gwynne proposed an alternative, arguably even simpler, formula that satisfies all known bounds \eqref{eq:boundsmallgamma}, \eqref{eq:dlower} and \eqref{eq:dupper},
\begin{equation}\label{eq:dinggwynne}
d_\gamma^{\mathrm{DG}} = 2 + \frac{\gamma^2}{2} + \frac{\gamma}{\sqrt{6}}.
\end{equation}
A quick glance at Figure \ref{fig:bounds}b leads one to conclude that it fits the numerical data at least as well as Watabiki's prediction does.

The goal of the current paper is to produce new numerical data to differentiate and hopefully rule out at least one of the predictions.
From Figure \ref{fig:bounds} it is clear that our chances are better when looking at smaller values of $\gamma$, where $d_\gamma^{\mathrm{W}}$ and $d_\gamma^{\mathrm{DG}}$ differ more substantially and the window between the upper and lower bound is bigger.
This is achieved by simulating a variety of random planar map models (Sections \ref{sec:planarmapmodels} and \ref{sec:fssplanarmaps}) as well as a discretized version of Liouville quantum gravity (Sections \ref{sec:lqg} and \ref{sec:fsslqg}).
The first study focuses on four models of random planar maps: Schnyder-wood-decorated triangulations ($\gamma=1$), bipolar-oriented triangulations ($\gamma=\sqrt{4/3}$), spanning-tree decorated quadrangulations ($\gamma=\sqrt{2}$) and uniform quadrangulations ($\gamma=\sqrt{8/3}$).
These models, the first two of which have not been studied numerically before, are particularly convenient because they can be simulated very efficiently, allowing high statistics to be obtained for surfaces with millions of vertices.
Moreover, their continuum limits in relation with Liouville quantum gravity are understood in considerable detail \cite{Kenyon2015,Miller2016,Gwynne2019b,Li2017,Ding2018}.
In the second half of the paper we perform a numerical investigation of Liouville quantum gravity on a regular lattice, allowing one in principle to tune $\gamma$ to any desired value.

\paragraph{Acknowledgments} We thank the Niels Bohr Institute, University of Copenhagen, for the use of their computing facilities.

\section{Four statistical models of random planar maps}\label{sec:planarmapmodels}

To describe the models we employ terminology that is customary in the mathematical literature on random surfaces.
A \emph{planar map} is a multi-graph, i.e.\ an unlabeled graph with multiple edges between pairs of vertices and loops allowed, together with an embedding in the 2-sphere, such that the edges are simple and disjoint except where they meet at vertices (see e.g.\ Figure \ref{fig:spanningtreealgorithm}(a)).
Any two planar maps that can be continuously deformed into each other are considered identical.
We always take planar maps to be \emph{rooted}, meaning that they are equipped with a distinguished oriented edge called the \emph{root edge}.
In this way we ensure that a planar map does not have any non-trivial automorphisms, which greatly simplifies counting and random sampling.
A region in the sphere delimited by $d$ edges is called a \emph{face of degree $d$} (where the edges that are incident to the face on both sides are double-counted).
We denote by $\mathcal{M}^{(d)}_n$ the set of $d$-angulations of size $n$, i.e.\ the set of planar maps with $n$ faces that are all of degree $d$.
In particular $\mathcal{M}^{(3)}_n$ is the set of \emph{triangulations} with $n$ triangles and $\mathcal{M}^{(4)}_n$ is the set of \emph{quadrangulations} with $n$ squares (of genus $0$ by construction).
Although we will not use this, it is convenient to think of a planar map as describing a piecewise-flat surface obtained by associating to each face of degree $d$ a regular $d$-gon of side length $1$ and performing the gluing of the polygons according to the incidence relations of the planar map (this underlies for instance the visualizations of the large planar maps in Figure \ref{fig:simulations3d}).

Since there are finitely many quadrangulations (or triangulations) of fixed size $n$, the simplest model of a random surface is to sample one such quadrangulation at random with equal probability, called the \emph{uniform quadrangulation} of size $n$ (see Section \ref{sec:uniformquad}).
To obtain different distributions one may couple the planar map to a variety of statistical systems.
In the cases at hand these systems consist of decorations of a planar map by a coloring of some of its edges subject to a number of constraints.
The systems are relatively simple in the sense that each decoration occurs with equal probability.
The number $Z^{*}(\map)$ of available decorations differs from one planar map $\map$ to another, which explains the effect of the statistical system on the geometry of the random surface.
In the language of statistical physics we are dealing with the \emph{canonical partition function}
\begin{equation}
Z^*_n = \sum_{\map\in \mathcal{M}_n^{(d)}} Z^*(\map) = \sum_{\map\in \mathcal{M}_n^{(d)}} \,\,\sum_{\text{decorations of }\map} 1. 
\end{equation}
From a combinatorial point of view it often hard to determine the number of decorations $Z^*(\map)$ of a given planar map $\map$, while the total number $Z_n^*$ of decorated planar maps is more easily accessible.
Similarly, from an algorithmic point of view, it is often much easier to generate a random decorated planar map (with uniform Boltzmann weight $1$) and then to forget the decoration, than it is to directly generate a random planar map with the non-trivial Boltzmann weight $Z^*(\map)$.
Below we describe four models fitting this bill and we describe in some detail the algorithms used to generate them.

In general the canonical partition function will asymptotically be of the form
\begin{equation}
Z_n^{*} \stackrel{n\to\infty}{\sim} C\, n^{\gamma_{\mathrm{s}}-2}\, \kappa^{n}
\end{equation}
for some constants $C$, $\kappa$ and $\gamma_{\mathrm{s}}$.
Whereas $C$ and $\kappa$ depend on the precise definition of the model, the \emph{string susceptibility} $\gamma_{\mathrm{s}}$ is universal, meaning that it typically only depends on the universality class of the system. 
It is related via the KPZ formula \cite{Knizhnik1988} to the central charge $c$ of the coupled statistical system,
\begin{equation}
\gamma_{\mathrm{s}} = \frac{c-1-\sqrt{(c-1)(c-25)}}{12}.
\end{equation}
In the continuum limit the geometry of a random surface coupled to a statistical system of central charge $c \in (-\infty,1]$ is believed to be described by Liouville quantum gravity with parameter $\gamma \in (0,2]$ related to $c$ and $\gamma_{\mathrm{s}}$ via
\begin{equation}\label{eq:cgamma}
c = 25 - 6 \left(\frac{2}{\gamma}+\frac{\gamma}{2}\right)^2, \quad \gamma_{\mathrm{s}} = 1 - \frac{4}{\gamma^2}.
\end{equation}
Table~\ref{tbl:couplings} summarizes the relevant values for the four models.
\begin{table}[h]
	\caption{\label{tbl:couplings}Critical exponents of the four models.}
	\begin{indented}
		\item[]
		\begin{tabular}{l  c c c}\hline
			& $\gamma_s$ & $c$ & $\gamma$ \\\hline
			Uniform quadrangulations {\bfseries(U)} & $-\frac{1}{2}$ & $0$ & $\sqrt{8/3}$ \\
			Spanning-tree-decorated quadrangulations {\bfseries(S)} & $-1$ & $-2$ & $\sqrt{2}$ \\
			Bipolar-oriented triangulations {\bfseries(B)} & $-2$ & $-7$ & $\sqrt{4/3}$ \\
			Schnyder-wood-decorated triangulations {\bfseries(W)} & $-3$ & $-\frac{25}{2}$ & $1$ \\\hline
		\end{tabular}
	\end{indented}
\end{table}

\subsection{Uniform quadrangulations {\bf (U)}}\label{sec:uniformquad}

As mentioned the simplest situation corresponds to quadrangulations with no decoration, i.e.\ to the uniform case $Z^{\mathrm{U}}(\map) = 1$ for $\map\in \mathcal{M}_n^{(d)}$.
The enumeration $Z_n^{\mathrm{U}}$ of (rooted) quadrangulations of size $U$ goes back to Tutte in the sixties \cite{Tutte1963} and is given explicitly by
\begin{equation}
Z_n^{\mathrm{U}} = \sum_{\map\in\mathcal{M}_n^{(4)}} 1 = 2\, \frac{(2n)!}{n!(n+2)!}\, 3^n \quad\stackrel{n\to\infty}{\sim}\quad \frac{2}{\sqrt{\pi}} n^{-5/2}\, 12^n.
\end{equation}
An efficient way to sample a quadrangulation of size $n$ uniformly at random uses the Cori-Vauquelin-Schaeffer bijection \cite{Schaeffer1998} (see e.g.\,\cite[Section 2.3]{Miermont2014} for a review), which provides a $2$-to-$1$ map between quadrangulations with an additional marked vertex and certain labeled trees.
Such trees can be generated easily via standard algorithms, after which the corresponding quadrangulations can be reconstructed.

\subsection{Spanning-tree-decorated quadrangulations {\bf (S)}}

\begin{figure}[h]
	\centering
	\includegraphics[width=\linewidth]{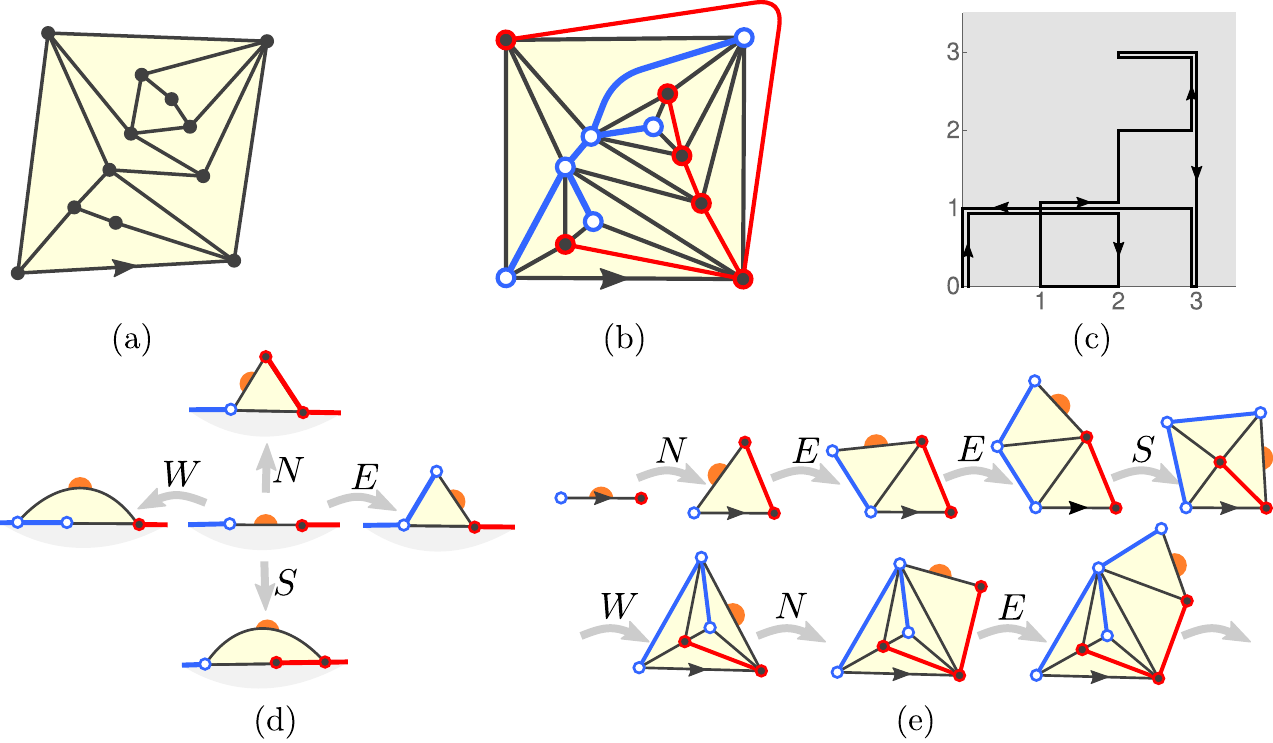}
	\caption{(a) A rooted quadrangulation in $\mathcal{M}^{(4)}_{10}$. (b) A decoration by spanning trees. (c) The corresponding $\{N,W,S,E\}$-walk in the quadrant: N-E-E-S-W-N-E-N-E-N-W-E-S-S-S-N-W-W-W-S. (d) The schematic operations corresponding to the steps of the walk. (e) The first seven steps in the construction. \label{fig:spanningtreealgorithm}}
\end{figure}

The first type of decorations we consider is that of spanning trees on a quadrangulation $\map\in\mathcal{M}_n^{(4)}$. 
Any quadrangulation admits a bipartition of its vertices, i.e.\ a black-white coloring of its vertices such that no two vertices of the same color are adjacent, that is unique if we specify that the origin of the root edge is colored white. 
A decoration of $\map$ amounts to a choice of diagonal in each face of $\map$ such that all diagonals combined form a pair of trees, one spanning the black vertices and the other spanning the white vertices (the red respectively blue tree in Figure \ref{fig:spanningtreealgorithm}b).
The exact enumeration also goes back to the sixties, in this case to Mullin \cite{Mullin1967}, and reads
\begin{equation}\label{eqn:ZS}
Z_n^{\mathrm{S}} = \sum_{\map\in\mathcal{M}_n^{(4)}} Z^{\mathrm{S}}(\map) = C_n C_{n+1}\quad\stackrel{n\to\infty}{\sim}\quad \frac{4}{\pi} n^{-3}\, 16^n.
\end{equation}
where $C_n = \frac{1}{n+1} \binom{2n}{n}$ is the $n$th Catalan number.
The quantity $C_nC_{n+1}$ also counts the number of two-dimensional lattice walks of length $2n$ with unit steps (denoted by the cardinal directions $\{N,W,S,E\}$) starting and ending at the origin and staying in the quadrant $\Z_{\geq 0}^2$ (Figure \ref{fig:spanningtreealgorithm}c).
This is explained by a natural encoding \cite{Mullin1967,Sheffield2016} of spanning-tree-decorated quadrangulations by such lattice walks.
Starting from a lattice walk the corresponding quadrangulation is constructed iteratively by starting with just the root edge with the left side designated \emph{active} (indicated in orange in Figure \ref{fig:spanningtreealgorithm}e) and performing the operations in Figure \ref{fig:spanningtreealgorithm}d consecutively for each step of the walk.

With the bijection in hand we can generate a random quadrangulation according to the partition function \eqref{eqn:ZS} from a uniform random $\{N,W,S,E\}$-walk in the quadrant of length $2n$. 
This can be achieved efficiently by using the decomposition
\begin{equation}
Z_n^{\mathrm{S}} = C_nC_{n+1} = \sum_{\ell=0}^{n}\binom{2n}{2\ell}C_\ell C_{n-\ell},
\end{equation}
where the summands count precisely the walks with $2\ell$ horizontal steps.
We may thus first sample $\ell$ with probability distribution $\binom{2n}{2\ell}C_\ell C_{n-\ell}/Z_n^{\mathrm{S}}$ and then randomly interleave two random Dyck paths of lengths $2\ell$ and $2n-2\ell$ (one for the horizontal and one for the vertical steps).

\subsection{Bipolar-oriented triangulations {\bf (B)} }

\begin{figure}[t]
	\centering
	\includegraphics[width=\linewidth]{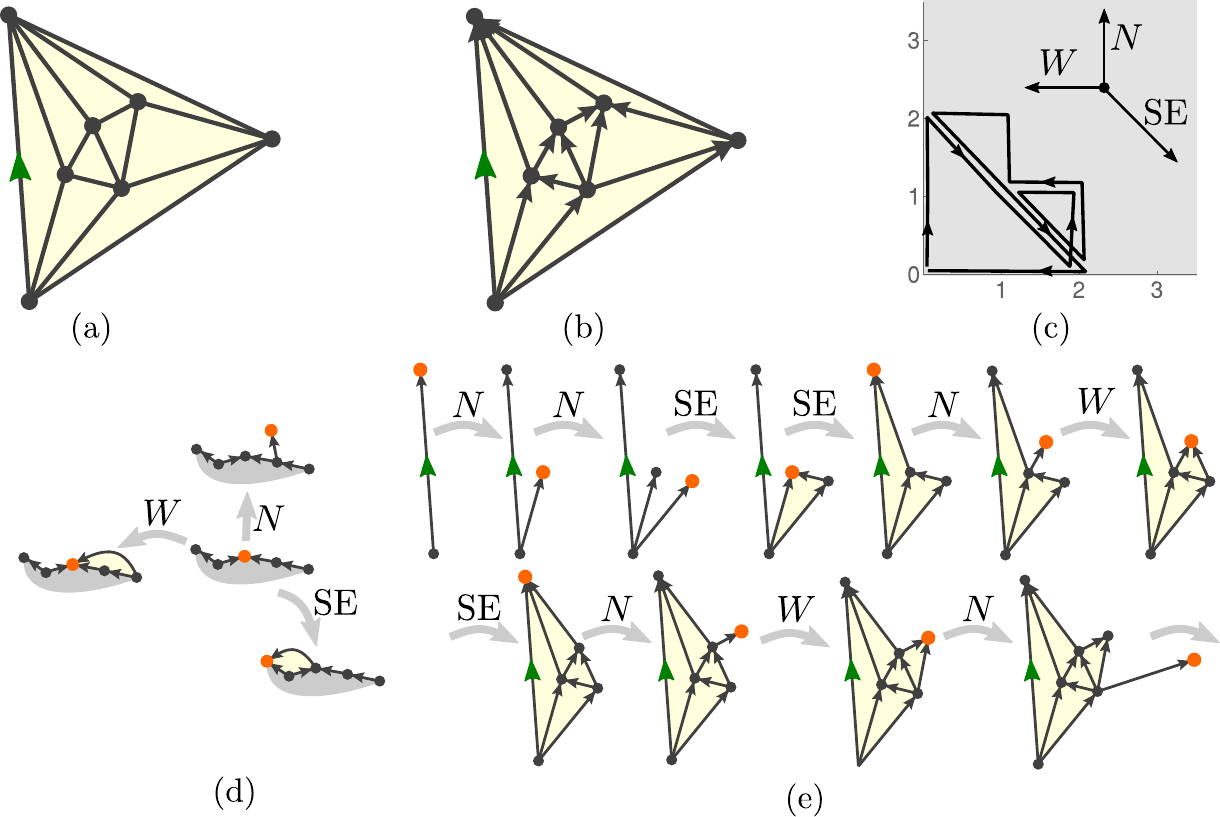}
	\caption{(a) A triangulation in $\mathcal{M}^{(3)}_{10}$ with the root in green. (b) A bipolar orientation (all edges oriented to the north). (c) The corresponding $\{N,W,SE\}$-walk in the quadrant: N-N-SE-SE-N-W-SE-N-W-N-W-SE-SE-W-W. (d) The schematic operations corresponding to the steps of the walk. (e) The first ten steps in the construction. \label{fig:bipolaralgorithm}}
\end{figure}
A bipolar orientation of a planar map $\map$ is an assignment of directions to each edge of $\map$ that is \emph{acyclic}, i.e.\ has no directed cycles, and possesses a single \emph{source} and \emph{sink}, i.e.\ a vertex with no incoming respectively outgoing edges. 
Here we consider triangulations $\map\in\mathcal{M}^{(3)}_n$ decorated with a bipolar orientation such that the origin and end-point of the root edge are respectively the source and the sink (Figure \ref{fig:bipolaralgorithm}b).
The total number of such bipolar-oriented triangulations with $2n$ triangles was also calculated by Tutte \cite[Equation (32)]{Tutte1973} (see also \cite[Proposition 5.3]{Bousquet-Melou2011}),
\begin{equation}
Z_{2n}^{\mathrm{B}} = \sum_{\map\in\mathcal{M}_{2n}^{(3)}} Z^{\mathrm{B}}(\map) = \frac{2(3n)!}{n!(n+1)!(n+2)!}\quad\stackrel{n\to\infty}{\sim}\quad \frac{\sqrt{3}}{\pi} n^{-4}\, 27^{n}.
\end{equation}
Recently an encoding by lattice walks has been discovered \cite{Kenyon2015} (see also \cite{Gwynne2016,Bousquet-Melou2019}) analogous to the spanning-tree-decorated quadrangulations.
The lattice walks again start and end at the origin and stay in the first quadrant, but now consist of $3n$ steps in $\{N,W,SE\}$ (Figure \ref{fig:bipolaralgorithm}c).
To construct the bipolar-oriented triangulation from the walk, one starts with just the root edge with its endpoint designated as the \emph{active} vertex (orange in Figure \ref{fig:bipolaralgorithm}e) and applies the operations in Figure \ref{fig:bipolaralgorithm}d according to the steps of the walk (except the very last).


To generate a $\{N,W,SE\}$-walk efficiently, we make use of the fact that the number of walks from $(x,y)$ to $(0,0)$ of length $3m+x+2y$ is known \cite[Proposition 9]{Bousquet-Melou2010} to be
\begin{equation}
\frac{(x+1)(y+1)(x+y+2)\,(3m+x+2y)!}{m!(m+y+1)!(m+x+y+2)!}.
\end{equation}
From this it follows that if a random $\{N,W,SE\}$-walk of length $3n$ is at $(x,y)$ after $3n-k$ steps, then the next step will be $N$, $W$ or $SE$ with probabilities
\begin{align}
&\text{N:}\quad\frac{(y+2)(x+y+3)(k-x-2y)}{3k(y+1)(x+y+2)}, \quad \text{W:}\quad\frac{x(x+y+1)(k+2x+y+6)}{3k(x+1)(x+y+2)}, \nonumber\\
&\text{SE:}\quad\frac{y(x+2)(k-x+y+3)}{3k(x+1)(y+1)}.
\end{align}
This allows one to sample the walk iteratively in quasi-linear time.

\subsection{Schnyder-wood-decorated triangulations {\bf (W)}}

\begin{figure}[t]
	\centering
	\includegraphics[width=\linewidth]{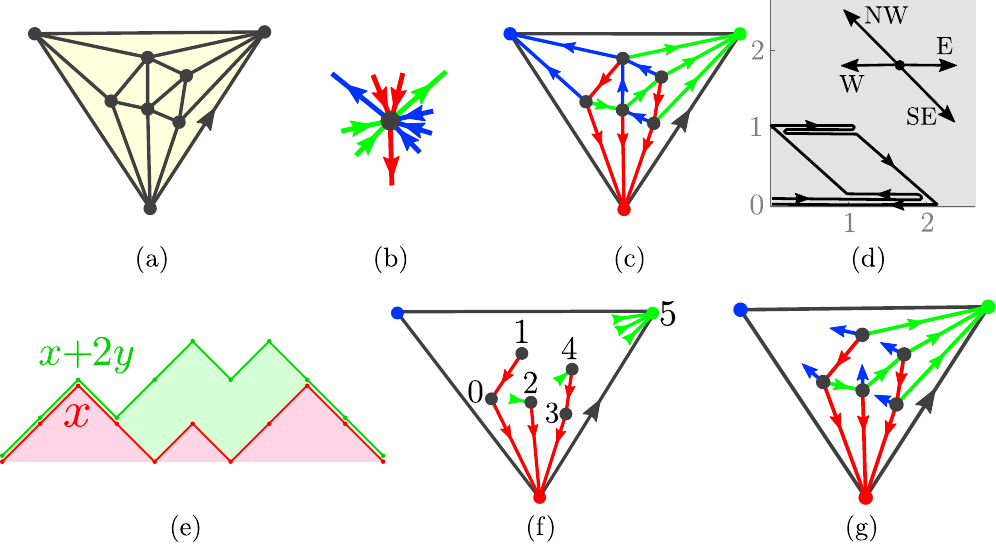}
	\caption{(a) A simple triangulation in $\mathcal{M}_{12}^{(3)}$. (b) The condition on the ordering of edges around inner vertices. (c) A Schnyder wood. (d) The corresponding $\{E,W,NW,SE\}$-walk of length $10$: E-E-W-NW-E-W-E-SE-W-W. (e) The corresponding double Dyck path. (f) The red tree together with the endpoints of the green edges as encoded by the double Dyck path. (g) As soon as the red and green edges are known, the blue edges (one for each uncolored vertex) are determined.   \label{fig:schnyderalgorithm}}
\end{figure}

Let $\map\in \mathcal{M}_{2n}^{(3)}$ be a \emph{simple} triangulation, meaning that it contains no double edges or loops (Figure \ref{fig:schnyderalgorithm}a).
Color the origin of the root edge, the endpoint of the root edge, and the remaining vertex incident to the triangle on the right of the root edge red, green, and blue respectively.
The edges that are incident to at least one uncolored vertex are called \emph{inner} edges.
A \emph{Schnyder wood} (also known as a \emph{realizer}) \cite{Schnyder1989} on $\map$ is a coloring in red, green, and blue and an orientation of all inner edges (Figure \ref{fig:schnyderalgorithm}c) satisfying the following properties:
\begin{itemize}
	\item Each uncolored vertex has precisely one outgoing edge of each color. Moreover, the incoming and outgoing edges of the various colors are ordered around the vertex as in Figure \ref{fig:schnyderalgorithm}b.
	\item The inner edges incident to a colored vertex are all incoming and of the same color as the vertex.
\end{itemize}
According to \cite[Corollary 19]{Bonichon2005} the number of Schnyder-wood-decorated triangulations with $2n$ triangles is
\begin{equation}
Z_{2n}^{\mathrm{W}} = \sum_{\map\in\mathcal{M}_{2n}^{(3)}} Z^{\mathrm{W}}(\map) =C_{n-1}C_{n+1} - C_n^2\quad\stackrel{n\to\infty}{\sim}\quad \frac{3}{2\pi} n^{-5}\, 16^{n}.
\end{equation}
Also for this model an encoding in terms of lattice walks in the quadrant is known \cite{Bonichon2005,Bernardi2009,Fusy2009,Li2017}, in this case consisting of $2n-2$ steps in $\{E,W,NW,SE\}$ starting and ending at the origin (Figure \ref{fig:schnyderalgorithm}d).
The way the encoding works is a bit different compared to the spanning-tree-decorated quadrangulations and bipolar-oriented triangulations.
First of all one represents the $\{E,W,NW,SE\}$-walk as a \emph{double Dyck path} of length $2n-2$, i.e.\ a pair of walks in the quadrant from the origin to $(2n-2,0)$ with steps in $\{NE,SE\}$ such that the first path does not go below the second (Figure \ref{fig:schnyderalgorithm}e).
This is achieved by plotting the graph of $x$ and $x+2y$ where $(x,y)$ ranges over the coordinates of the $\{E,W,NW,SE\}$-walk.  
To construct the triangulation we start with a single triangle with colored vertices and attach a red tree to the red vertex as encoded by the lower Dyck path in the usual fashion (``apply glue to the bottom of the Dyck path and squash horizontally'').
We then label the uncolored vertices of the red tree from $0$ to $n-2$ in the order in which they are encountered when tracing the contour of the tree in clockwise direction, and assign label $n-1$ to the green vertex (Figure \ref{fig:schnyderalgorithm}f).
The upper Dyck path is then used to determine the position of the endpoints of the green edges: for each $SE$-step that is preceded by a total of $k$ $NE$-steps we add an endpoint to the vertex with label $k$.
Since a single green edge has to start at each uncolored vertex (and end at the indicated positions), it is easy to see that there is a unique way to draw them while satisfying the condition in Figure \ref{fig:schnyderalgorithm}b. 
As soon as the red and green edges are drawn (Figure \ref{fig:schnyderalgorithm}g) the blue edges are also uniquely determined by this condition. 

As in the case of the bipolar-oriented triangulations, there is an efficient method to generate random $\{E,W,NW,SE\}$-walks of length $2n-2$.
The total number of such walks of length $2m+x$ starting at $(x,y)$ and ending at the origin is \cite[Proposition 11]{Bousquet-Melou2011}
\begin{equation}
\frac{(x+1)(y+1)(x+y+2)(x+2y+3)}{(2m+x+1)(2m+x+2)(2m+x+3)^2}\binom{2m+x+3}{m-y}\binom{2m+x+3}{m+1}.
\end{equation}
It follows that if a random $\{E,W,NW,SE\}$-walk of length $2n-2$ is at $(x,y)$ after $2n-2-k$ steps that the next step will be $E,W,NW,SE$ with probabilities
\begin{align*}
\text{E:}&\quad\frac{(x+2) (k-x+2) (x+y+3) (x+2 y+4) (k-x-2 y)}{4 k (k+2) (x+1) (x+y+2) (x+2y+3)},\\
\text{W:}&\quad\frac{x (k+x+4) (x+y+1) (x+2 y+2) (k+x+2 y+6)}{4 k (k+2) (x+1) (x+y+2) (x+2y+3)},\\
\text{NW:}&\quad\frac{x (y+2) (k+x+4) (x+2 y+4) (k-x-2 y)}{4 k (k+2) (x+1) (y+1) (x+2y+3)},\\
\text{SE:}&\quad\frac{(x+2)y (k-x+2) (x+2 y+2) (k+x+2 y+6)}{4 k (k+2) (x+1) (y+1) (x+2 y+3)}.
\end{align*}

\section{Finite-size scaling analysis of (dual) graph distances}\label{sec:fssplanarmaps}

As discussed in the introduction, the Hausdorff dimension $d_\gamma$ for $\gamma = \sqrt{8/3},\sqrt{2},\sqrt{4/3},1$ agrees with the growth exponent of the volume $|\mathrm{Ball}_r(\map_n)|$ of the ball of radius $r$, i.e.\ the number of vertices that have graph distance at most $r$ from a random initial vertex, in a random map $\map_n$ of size $n$ sampled from model (U), (S), (B), (W) respectively,\footnote{In probabilistic terms, the limit can be understood as a limit in distribution as $n\to\infty$ followed by an almost sure limit as $n\to\infty$ \cite[Theorem 1.6]{Ding2018}.}
\begin{equation}\label{eq:growth}
\frac{\log |\mathrm{Ball}_r(\map_n)|}{\log r} \xrightarrow[\substack{n,r\to\infty \\n\gg r}]{} d_\gamma.
\end{equation} 
Since we cannot attain the limit $n,r\to\infty$ in simulations, we employ the finite-size scaling method to estimate the exponents.
To this end, we need to make a few additional, but reasonable, assumptions.
Let $R_n$ be the graph distance between two uniformly sampled vertices in a random planar map of size $n$ (sampled from one of the models).
For integer $r$ we set $\rho_n^{(*)}(r) = \mathbb{P}(R_n = r)$ to be the probability that this distance is $r$, and we extend $\rho_n^{(*)}$ to a continuous function $\rho_n^{(*)} : (0,\infty)\to[0,\infty)$ by interpolation. 
We assume that for any $x>0$
\begin{equation}\label{eq:scalingassumption}
\lim_{n\to\infty} n^{1/d_\gamma}\rho_n^{(*)}(x \,n^{1/d_\gamma}) = \rho^{(*)}(x)
\end{equation}
for a continuous probability distribution $\rho^{(*)}$ on $(0,\infty)$ that depends only on the model $(\ast)$.
This is slightly stronger than the requirement that $R_n / n^{1/d_\gamma}$ converges in distribution as $n\to\infty$. 
As we will see shortly \eqref{eq:scalingassumption} is well supported by our data and known to be correct for uniform quadrangulations (with an explicit limit \cite{Bouttier2003}).\footnote{In the case of spanning-tree decorated quadrangulations \cite[Theorem 1.4]{Gwynne2019b} comes close by identifying the scaling of the diameter of $\map_n$ with growing $n$.}
Note, however, that it does not quite imply \eqref{eq:growth}, nor is it implied by \eqref{eq:growth}.

\begin{figure}[t]
	\centering
	\includegraphics[width=.6\linewidth]{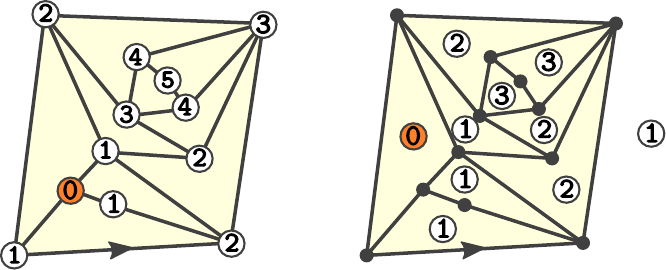}
	\caption{The graph distance (left) and dual graph distance (right) in a quadrangulation.\label{fig:graphdistance}}
\end{figure}

We estimate $\rho_n^{(*)}$ with high accuracy for each model $(\ast)$ and sizes $n$ ranging from $2^8$ up to $2^{24}$ ($\approx 17$ million) by sampling a large ensemble of random planar maps ($10^7$ for small $n$ and $10^5$ for $n\geq 2^{19}$).
In each random planar map we pick a single uniform random vertex and determine the graph distance to all other vertices in the map (Figure~\ref{fig:graphdistance}).
All these distances for the planar maps in an ensemble are included in a histogram, which upon normalization and interpolation provides our estimate of $\rho_n^{(*)}$.
It turns out to be convenient to supplement the analysis with another distribution $\rho_n^{(*)\dagger}$ that relies on a different notion of distance: the \emph{dual} graph distance between two uniformly sampled faces in the planar map (right side of Figure \ref{fig:graphdistance}).
It is estimated in an analogous way, this time picking a uniform random face and determining the distances to all other faces.

To test the convergence \eqref{eq:scalingassumption} we choose optimal parameters $k_n$ to ``collapse'' the functions $x \mapsto k_n^{-1}\,\rho_n^{(*)}(k_n^{-1}x)$.
To be precise, we denote by $n_0 = 2^{24}$ the largest system size and take $\rho^{(*)}_{n_0}$ as the reference distribution.
Then for each $n \leq n_0$, $k_n$ is obtained by fitting $x \mapsto k_n^{-1}\,\rho_n^{(*)}(k_n^{-1}x)$ to $x \mapsto \rho_{n_0}^{(*)}(x)$, such that $k_n > 1$ for $n<n_0$ and $k_{n_0}=1$ by construction.
In the fit we choose to only take into account the portion of the histogram for which $\rho_n^{(*)}(r) \geq \frac{1}{5} \max_{r'} \rho_n^{(*)}(r')$, thus avoiding the tails of the distribution that are more prone to discretization effects. 

\begin{figure}
	\centering
	\includegraphics[width=\linewidth]{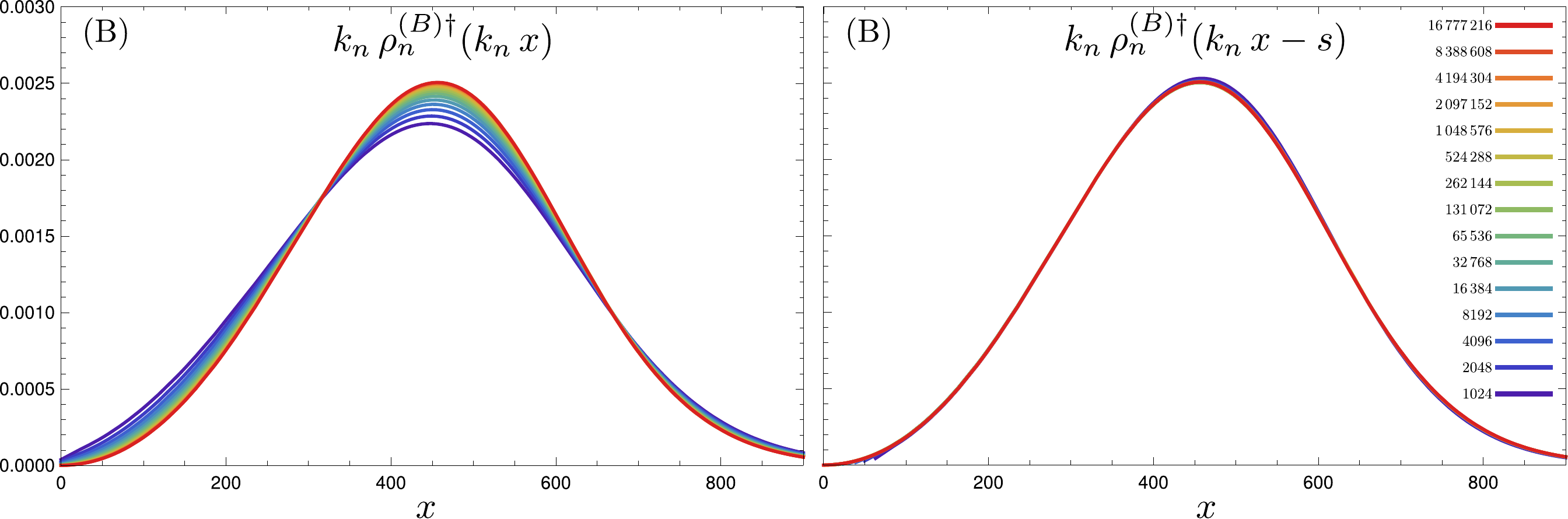}
	\caption{Finite-size scaling for the dual graph distance in bipolar-oriented triangulations without shift (left) and with shift $s=2.941$ (right). The effect in the case of graph distance is less pronounced. \label{fig:collapseshift}}
\end{figure}

The collapse of $\rho_n^{(B)\dagger}$ in the case of dual graph distance on bipolar-oriented triangulations is shown in the left plot of Figure \ref{fig:collapseshift}.
A qualitative convergence is certainly observed, but one has to go to quite large sizes for the curves to become indistinguishable.
A common technique \cite{Ferrenberg1991,Ambjorn1995,Ambjorn1998} to improve the collapse is by introducing a \emph{shift} $\rho_n^{(*)}(r) \to \rho_n^{(*)}(r-s)$ in the histograms before performing the scaling, where $s\in\R$ is independent of $n$.
For any fixed $s$, the convergence $n^{1/d_\gamma} \rho_n^{(*)}(x n^{1/d_\gamma} - s) \xrightarrow{n\to\infty} \rho^{(*)}(x)$ is of course equivalent to our scaling ansatz \eqref{eq:scalingassumption}.
One may think of this shift as absorbing a subleading correction in \eqref{eq:scalingassumption} or, if you like, accounting for the freedom we have in the discrete setting to assign distance $s$ instead of $0$ to the initial vertex/face.
The optimal shift $s$ is determined by fitting $x\mapsto k_n^{-1} \rho_n^{(*)}(k_n^{-1} (x+s_n) - s_n)$ to $x \mapsto \rho_{n_0}^{(*)}(x)$ for each $n$ and taking $s$ to be a (weighted) average of the values $s_n$.
This way we fix $s$ once and for all to the values in Table~\ref{tbl:shift}.

\begin{table}[h]
	\caption{\label{tbl:shift}Optimal shift parameters}
	\begin{indented}
		\item[]
		\begin{tabular}{ccc}\hline
			Model & Graph distance & Dual graph distance \\\hline
			(U) & $s=0.940$ & $s=4.608$ \\
			(S) & $s=0.557$ & $s=3.019$ \\
			(B) & $s=0.359$ & $s=2.941$ \\
			(W) & $s=0.439$ & $s=2.629$ \\\hline
		\end{tabular}
	\end{indented}
\end{table}

\begin{figure}[t]
	\centering
	\includegraphics[width=\linewidth]{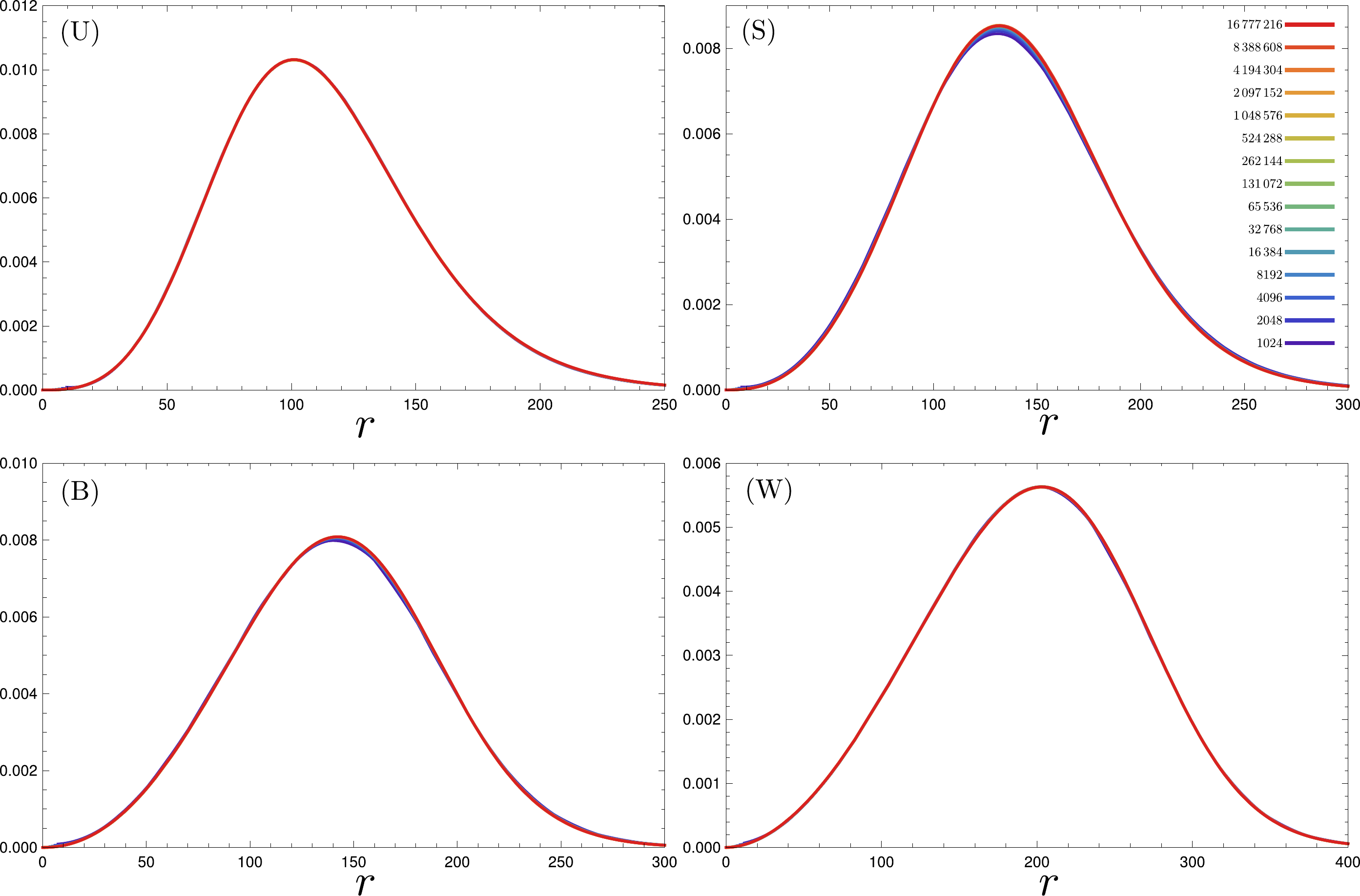}
	\caption{Finite-size scaling of the graph distance in the four models with sizes ranging from $2^{10}=1024$ to $2^{24}=16\,777\,216$ and shifts as listed in the table. \label{fig:fss} }
\end{figure}

With $s$ fixed we determine the optimal scaling $k_n$ yet again by fitting $x\mapsto k_n^{-1} \rho_n^{(*)}(k_n^{-1} (x+s) - s)$ to $x \mapsto \rho_{n_0}^{(*)}(x)$.
The result in the case of $\rho_n^{(B)\dagger}$ is shown in the right plot of Figure \ref{fig:collapseshift}.
This time all curves for $n\gtrsim 2^{13}$ become indistinguishable at the resolution of the plot.
The finite-size scaling of the graph distance including the shift is shown in Figure \ref{fig:fss} for all four models.
The very accurate scaling lends support to the existence of a continuous probability distribution $\rho^{(*)}$ in the limit \eqref{eq:scalingassumption}.

\begin{figure}[t]
	\centering
	\includegraphics[width=\linewidth]{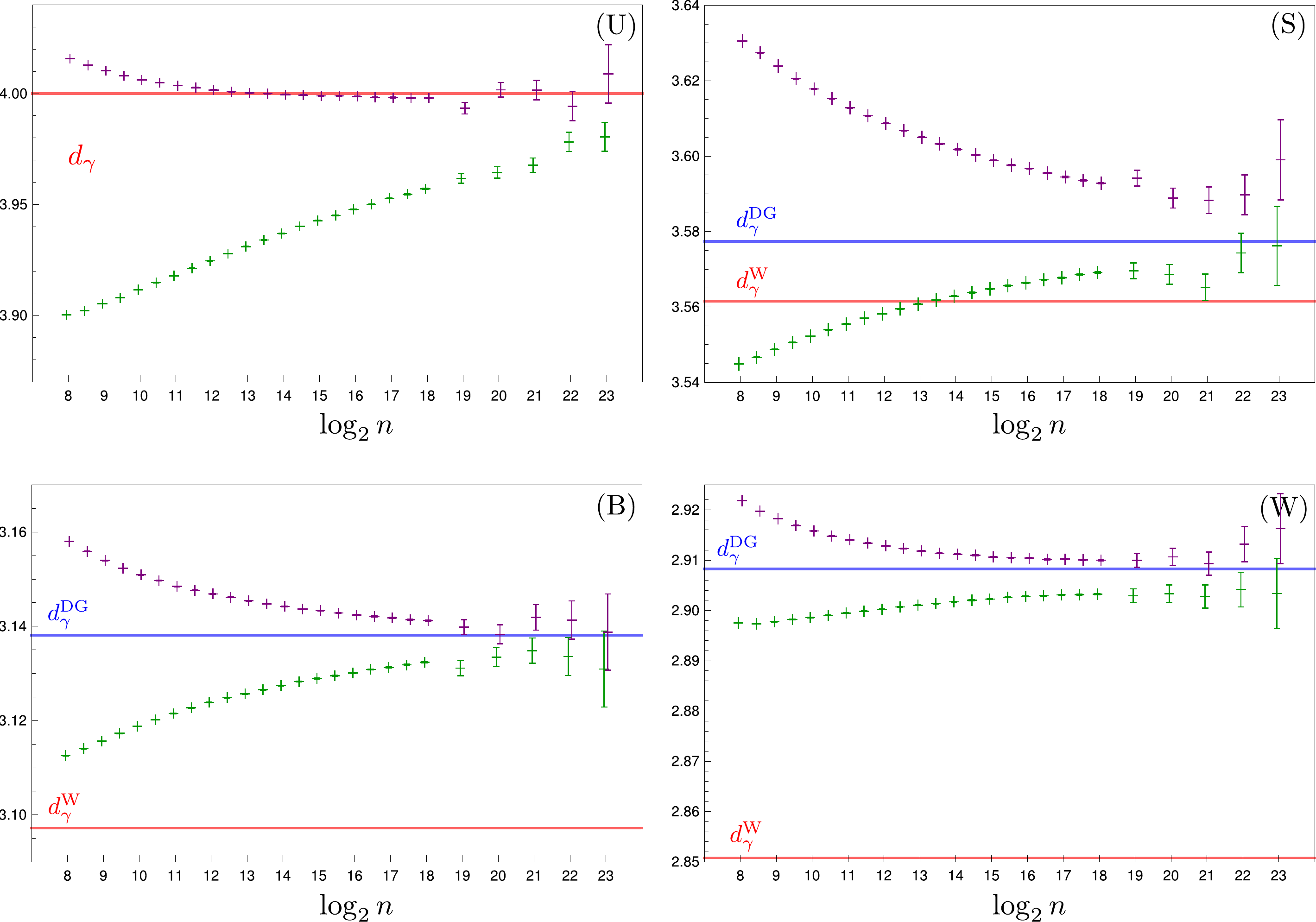}
	\caption{Plots of $\frac{\log(n_{0}/n)}{\log(k_{n}/k_{n_{0}})}$ with statistical error bars for all four models, showing both graph distance (purple) and dual graph distance (green) measurements. Watabiki's and Ding-Gwynne's predictions are indicated by horizontal lines. \label{fig:ratioplot}}
\end{figure}

If \eqref{eq:scalingassumption} is satisfied then 
\begin{equation}\label{eq:knasymptotics}
k_n \,\stackrel{n\to\infty}{\sim}\, c\, (n/n_0)^{-1/d_\gamma}
\end{equation}
for some constant $c \approx 1$.
To get a first idea of the rate of convergence to the asymptotics \eqref{eq:knasymptotics}, we plot in Figure \ref{fig:ratioplot} the logarithmic ratios
\begin{equation}
\frac{\log(n_{0}/n)}{\log(k_{n}/k_{n_{0}})}
\end{equation}
with statistical error bars for the four models using both the graph distance (purple) and its dual (green).
The advantage of considering the different distance measurements becomes clear upon inspecting these plots: the deviations from the scaling relation \eqref{eq:knasymptotics} appear with different sign, allowing one in principle to estimate the exponent $d_\gamma$ by eye at the point where the two curves converge. 
It is also immediately clear that the data is incompatible with $d_\gamma^{\mathrm{W}}$ in the case of bipolar-oriented and Schnyder-wood-decorated triangulations, and is much closer to $d_\gamma^{\mathrm{DG}}$.

\begin{figure}
	\centering
	\includegraphics[width=\linewidth]{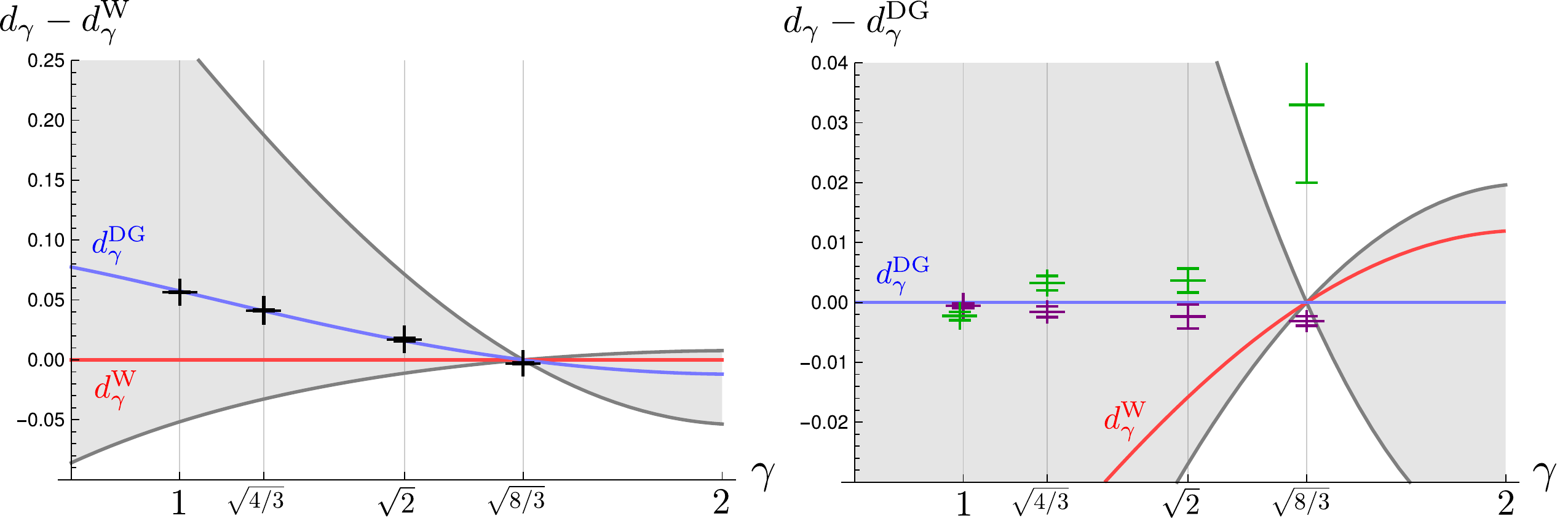}
	\caption{The plot on the left shows the estimates of $d_\gamma$ in comparison to $d_\gamma^{\mathrm{W}}$ in the same scale as Figure \ref{fig:bounds}b. The right plot shows the individual estimates of $d_\gamma$ using the graph distance (purple) and dual graph distance (green) in comparison with $d_\gamma^{\mathrm{DG}}$. \label{fig:result} }
\end{figure}

To accurately estimate $d_\gamma$, we make an ansatz for the leading-order correction to \eqref{eq:knasymptotics} of the form
\begin{equation}\label{eqn:kfitleadingorder}
k_n \approx \left(\frac{n}{n_0}\right)^{-\frac{1}{d}} \left(a + b\left(\frac{n}{n_0}\right)^{-\delta}\right),
\end{equation}
where $a$ is close to $1$, $b$ is small and $\delta > 0$. 
The best fits are given in Table~\ref{tbl:correctionfit}, including the statistical errors on $d$.
Finally, combining the data from both distance measures yields the estimates for the Hausdorff dimension recorded in Table~\ref{tbl:pmdimensions} and plotted in Figure~\ref{fig:result}.

\begin{figure}[h]
	\centering
	\includegraphics[width=\linewidth]{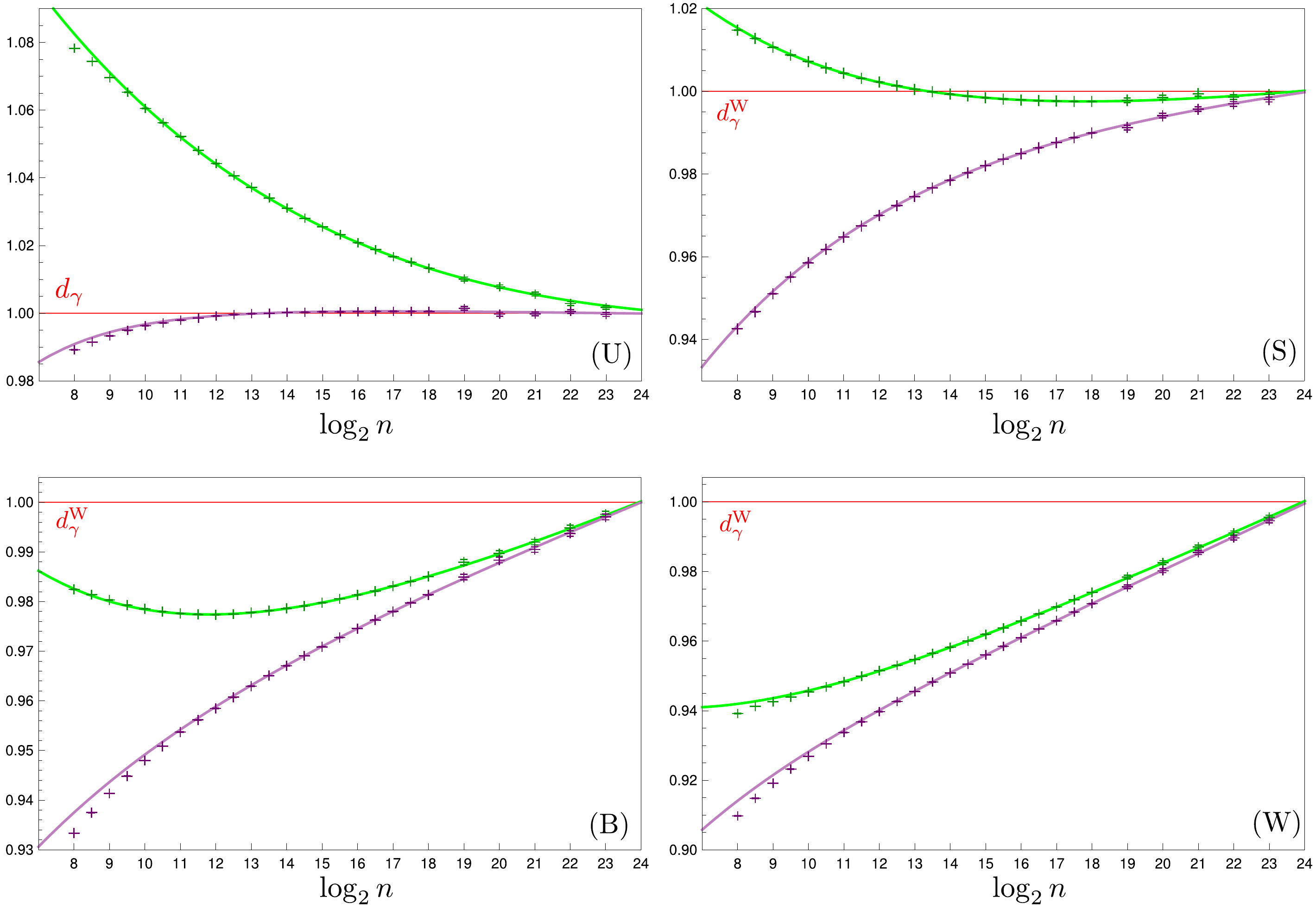}
	\caption{Plots of $k_n (n/n_0)^{1/d^{\mathrm{W}}_\gamma}$ with the scaling parameters $k_n$ established via finite-size scaling of the graph distance (purple) and dual graph distance (green). The solid curves correspond to best fits of the ansatz \eqref{eqn:kfitleadingorder}. \label{fig:correctionplot} }
\end{figure}

\begin{table}[h]
	\caption{\label{tbl:correctionfit}Parameters of the best fit of the data to the ansatz \eqref{eqn:kfitleadingorder}.}
	\begin{indented}
		\item[]
		\begin{tabular}{ccccc}\hline
			Model & $d$ & $\delta$ & $a$ & $b$ \\\hline
			(U) & $3.9969 \pm 0.0013$ & $0.57$ & $0.99992$ & $-0.000020$ \\
			(U)$\dagger$ & $4.037 \pm 0.024$ & $0.16$ & $0.9795$ & $0.0206$ \\
			(S) & $3.575 \pm 0.006$ & $0.26$ & $1.0025$ & $-0.00275$ \\
			(S)$\dagger$ & $3.581 \pm 0.004$ & $0.26$ & $0.9982$ & $0.00190$\\
			(B) & $3.136 \pm 0.002$ & $0.29$ & $1.00081$ & $-0.00084$ \\
			(B)$\dagger$ & $3.141 \pm 0.003$ & $0.27$ & $0.9984$ & $0.0018$ \\
			(W) & $2.9077 \pm 0.0010$ & $0.41$ & $0.99968$ & $-0.00014$ \\
			(W)$\dagger$ & $2.906 \pm 0.002$ & $0.33$ & $0.99985$ & $0.00038$ \\\hline
		\end{tabular}
	\end{indented}
\end{table}

\begin{table}[h]
	\caption{\label{tbl:pmdimensions}Hausdorff dimension estimates for the four planar map models.}
	\begin{indented}
		\item[]
	\begin{tabular}{ccccc}\hline
		Model & $\gamma$ & $d_\gamma$ & $d_\gamma^{\mathrm{W}}$ & $d_\gamma^{\mathrm{DG}}$ \\\hline
		(U) & $\sqrt{8/3}$ & $3.9970\pm 0.0013$ & 4.0000 & 4.0000 \\
		(S) & $\sqrt{2}$ & $3.5791\pm 0.0033$ & 3.5616 & 3.5774 \\
		(B) & $\sqrt{4/3}$ & $3.1375\pm 0.0017$ & 3.0972 & 3.1381 \\
		(W) & 1 & $2.9074\pm 0.0009$ & 2.8508 & 2.9083 \\\hline
	\end{tabular}
	\end{indented}
\end{table}

\FloatBarrier
\section{Hausdorff dimensions in Liouville quantum gravity}\label{sec:lqg}

Simulations of the four planar maps models have provided accurate estimates of the Hausdorff dimension $d_\gamma$ for $\gamma = 1,\sqrt{4/3},\sqrt{2},\sqrt{8/3}$. 
In principle one can extend this method by exploring new discrete models that live in other universality classes.
However, finding such models that allow for efficient simulation is a non-trivial task\footnote{The \emph{mated-CRT maps} of \cite{Gwynne2017a} are a good candidate for arbitrary $\gamma\in(0,2)$.}.
An alternative route is to start from the continuum description of Liouville quantum gravity.

Formally one thinks of the random two-dimensional metric as a Weyl-transformation $g_{ab}=e^{\gamma \phi}\hat{g}_{ab}$ of a fixed background metric $\hat{g}_{ab}$ on the surface.
The field $\phi$ is sampled with probability proportional to $e^{-S_{\mathrm{L}}[\phi]}$ with the Liouville action given by \cite{Knizhnik1988,David1988,Distler1989}
\begin{equation}\label{eq:liouvilleaction}
S_{\mathrm{L}}[\phi] = \frac{1}{4\pi} \int \rmd^2 x\sqrt{\hat{g}(x)}(\hat{g}^{ab}\partial_a\phi\partial_b\phi + Q \hat{R}\phi+ 4\pi\lambda e^{\gamma\phi}), 
\end{equation}
where $\hat{R}$ is the scalar curvature of $\hat{g}_{ab}$, $\lambda$ the \emph{cosmological constant}, and $Q = 2/\gamma + \gamma/2$.
Since we are only interested in the local properties of the metric $g_{ab}$ we may as well choose $\lambda=0$ and fix $\hat{g}_{ab}=\delta_{ab}$ to the standard Euclidean metric on the unit torus (using periodic coordinates $x \in \R^2/\Z^2$).
In this case the action \eqref{eq:liouvilleaction} becomes that of a scalar free field, 
\begin{equation}\label{eq:gffaction}
S_{\mathrm{L}}[\phi]=S_{\mathrm{GFF}}[\phi] = \frac{1}{4\pi} \int_{\R^2/\Z^2} \rmd^2 x\, \nabla\phi\cdot\nabla\phi.
\end{equation}
The random field $\phi$ sampled with (suitably regularized) probability $e^{-S_{\mathrm{GFF}[\phi]}}$ is called the \emph{Gaussian free field} in the mathematical literature \cite{Sheffield2007}.
The pointwise values of $\phi$ are not well-defined, but the field can be rigorously understood as a random generalized function (or distribution) living in an appropriate Sobolev space.
This implies that the identification $g_{ab} = e^{\gamma \phi(x)}\hat{g}_{ab}$ cannot make literal sense as a random Riemannian metric without choosing a regularization scheme.
At the level of the volume form $\sqrt{g}\rmd^2 x = e^{\gamma\phi(x)}\rmd^2x$ this can achieved by imposing an ultraviolet cutoff $\epsilon$ on $\phi$, e.g. by setting $\phi_\epsilon(x)$ to be the average of $\phi$ over a circle of radius $\epsilon$ centered at $x$, and considering the limit
\begin{equation}\label{eq:liouvillemeasure}
\lim_{\epsilon\to 0} \epsilon^{\gamma^2/2} e^{\gamma\phi_\epsilon(x)} \rmd^2x
\end{equation}
viewed as a random measure, called the \emph{$\gamma$-Liouville measure} \cite{Duplantier2011}.

Determining a regularization scheme of $g_{ab} = e^{\gamma \phi(x)}\hat{g}_{ab}$ that gives rise to well-defined geodesic distances $d(x,y)$ between pairs of points is more challenging.
The intuitive reason for this is that the distance between two points is realized by a curve that generically has a fractal structure (in the Euclidean background metric), meaning that its length will be quite sensitive to the way the ultraviolet cutoff is imposed.
Nevertheless, there has been much progress in recent years, resulting in at least two different approaches.

\subsection{Liouville graph distance}

The \emph{Liouville graph distance} $D_{\gamma,\delta}(x,y)$ between two points is defined as the fewest number of Euclidean disks of arbitrary radius, but volume at most $\delta$ as measured by the $\gamma$-Liouville measure, needed to cover a path connecting $x$ and $y$ \cite{Ding2018a,Ding2019,Ding2018}.
This definition should be viewed as the analogue of the (dual) graph distance in a random planar map of size $n\approx \delta^{-1}$, where the distance is the fewest number of faces (which all have equal volume $\approx 1$) one has to traverse to get from one vertex to another.
It has been shown rigorously \cite[Theorem 1.4]{Ding2018} that $D_{\gamma,\delta}(x,y)$ for fixed $x$ and $y$ is of order $\delta^{-1/d_\gamma}$, i.e.
\begin{equation}
\lim_{\delta\to 0} \frac{\log D_{\gamma,\delta}(x,y)}{\log \delta} = - \frac{1}{d_\gamma}.
\end{equation}
This provides one avenue to measure $d_\gamma$ numerically, as was done by Ambj\o rn and the second author in \cite{Ambjorn2014}.
There a discrete $\gamma$-Liouville measure was constructed by exponentiating a discrete Gaussian free field (see Section \ref{sec:dlfpp} below) on a $w\times w$ square lattice with periodic boundary conditions.
Instead of finding paths of disks of volume $\delta$ connecting pairs of points, distances were obtained from a Riemannian metric with local density constructed from averaging the $\gamma$-Liouville measure over such disks of volume $\delta$, for which one expects similar behaviour.
Estimates on $d_\gamma$ obtained in \cite{Ambjorn2014} are shown in green in Figure \ref{fig:bounds}b.

\subsection{Liouville first passage percolation}
Following \cite{Benjamini2010,Ding2018a,Ding2019,Ding2018,Dubedat2019}, the \emph{Liouville first passage percolation} distance $D_{\xi,\epsilon}(x,y)$ between two points $x$ and $y$ is given for $\xi>0$ in terms of the regularized Gaussian free field $\phi_\epsilon$ by
\begin{equation}\label{eq:fpp}
D_{\xi,\epsilon}(x,y) = \inf_{\Gamma:x\to y} \int_0^1 e^{\xi \phi_\epsilon(\Gamma(t))} |\Gamma'(t)|\rmd t, 
\end{equation}
where the infimum is over piecewise differentiable paths $\Gamma$ with $\Gamma(0)=x$ and $\Gamma(1)=y$.

In the hope to construct a metric for $\gamma$-Liouville quantum gravity one should \emph{not} take $\xi = \gamma / 2$, which would arise from naively regularizing the Riemannian metric as $e^{\gamma \phi_\epsilon(x)} \hat{g}_{ab}$.
Assuming the existence of a Hausdorff dimension $d_\gamma$, one would like an overall scaling of the volume (as measured by the $\gamma$-Liouville measure) by a factor $C$ to amount to a scaling of the geodesic distances by $C^{1/d_\gamma}$.
The former is achieved by a constant shift $\phi(x) \to \phi(x) + \frac{1}{\gamma}\log C$ in \eqref{eq:liouvillemeasure}, leading to an overall scaling of \eqref{eq:fpp} by $C^{\xi / \gamma}$, hence suggesting the relation
\begin{equation}\label{eq:xi}
\xi = \gamma / d_\gamma.
\end{equation}
On the other hand, under a coordinate transformation $x\mapsto x' = C x$ one should transform $\phi(x)\mapsto\phi'(x') = \phi(x) - Q \log C$ with $Q=2/\gamma+\gamma/2$ in order to preserve the $\gamma$-Liouville measure \eqref{eq:liouvillemeasure} \cite{Duplantier2011}.
Accordingly, \eqref{eq:fpp} transforms as
\begin{align}
D'_{\xi,\epsilon}(x',y') &= \inf_{\Gamma:x\to y} \int_0^1 e^{\xi \phi_\epsilon(C\Gamma(t))-\xi Q\log C} |C\Gamma'(t)|\rmd t\\
&= C^{1-\xi Q}\inf_{\Gamma:x\to y} \int_0^1 e^{\xi \phi_{\epsilon/C}(\Gamma(t))} |\Gamma'(t)|\rmd t = C^{1-\xi Q} D_{\xi,\epsilon/C}(x,y).
\end{align}
Equality in the limit $\epsilon \to 0$ can only be achieved if $D_{\xi,\epsilon}(x,y)$ scales as $\epsilon^{1-\xi Q}$ as $\epsilon\to 0$ (see \cite[Section 2.3]{Ding2018} for a similar heuristic).
Indeed, it was proven rigorously in \cite[Theorem 1.5]{Ding2018} that the following limit holds (in probability)
\begin{equation}\label{eq:lambda}
\frac{\log D_{\xi,\epsilon}(x,y)}{\log \epsilon} \xrightarrow{\epsilon\to 0} \lambda(\xi),\quad \lambda(\xi) = 1-\xi Q = 1 - \frac{2}{d_\gamma} - \frac{\gamma^2}{2 d_\gamma}.
\end{equation}
See \cite[Theorem 1.1]{Gwynne2019a} for results on the limit of $\epsilon^{-\lambda}D_{\xi,\epsilon}(x,y)$ as a metric space.

Note that if we can determine $\lambda(\xi)$ for some value of $\xi$ satisfying $1-\lambda(\xi) > 2\xi$, then \eqref{eq:xi} and \eqref{eq:lambda} can be inverted to determine a pair of values $\gamma$ and $d_\gamma$.
This provides a different route towards numerical estimates of $d_\gamma$. 
Watabiki's formula \eqref{eq:watabiki} and Ding \& Gwynne's formula \eqref{eq:dinggwynne} correspond to the particularly simple relations
\begin{equation}
\lambda^{\mathrm{W}}(\xi) = \xi^2, \qquad \lambda^{\mathrm{DG}}(\xi) = \frac{\xi}{\sqrt{6}}.
\end{equation}

\subsection{Discrete Liouville first passage percolation (DLFPP)}\label{sec:dlfpp}

\begin{figure}[t]
	\captionsetup[subfigure]{labelformat=empty}
	\centering
	\subfloat[$\xi=0.1$]{\includegraphics[width=.3\linewidth]{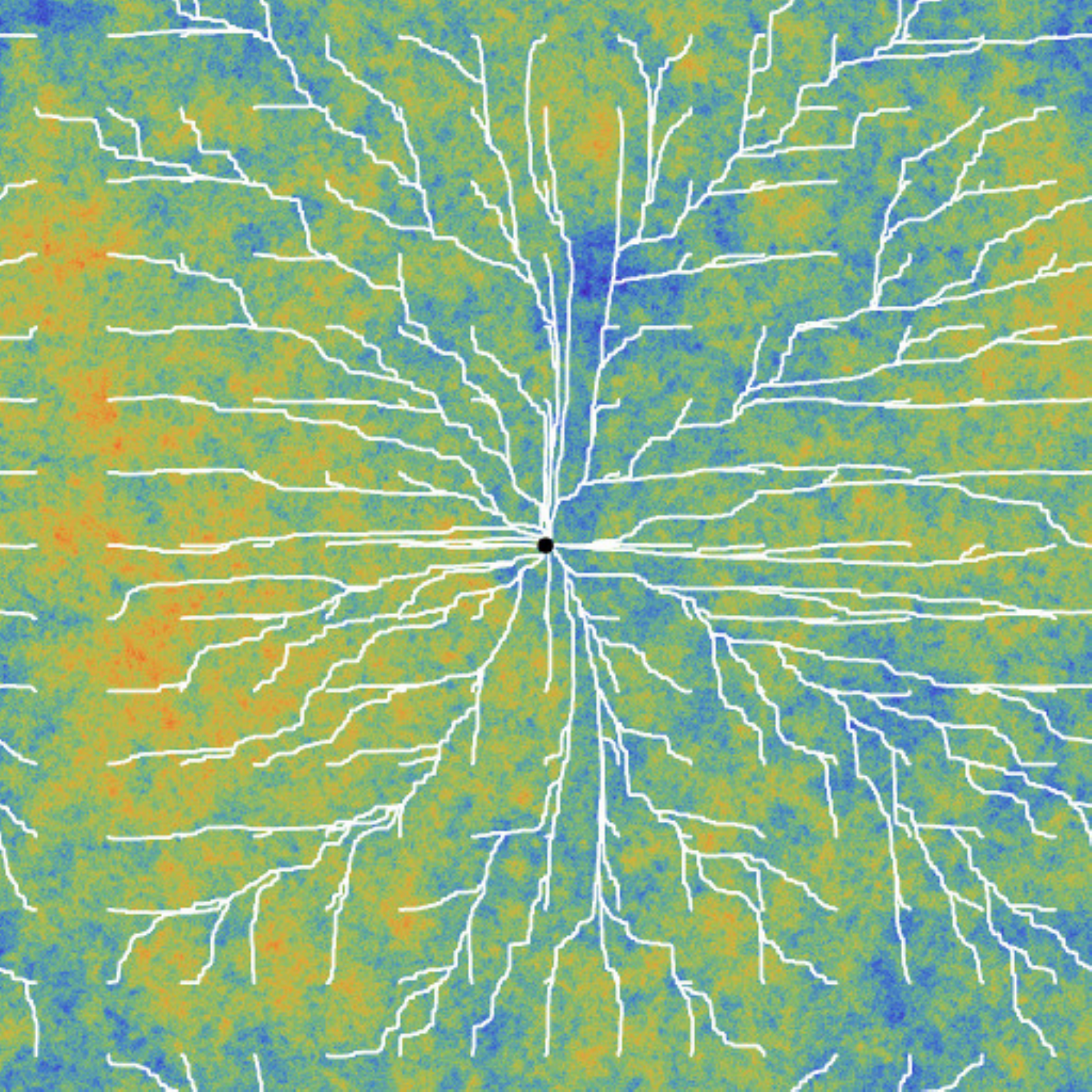}}\hfill
	\subfloat[$\xi=0.25$]{\includegraphics[width=.3\linewidth]{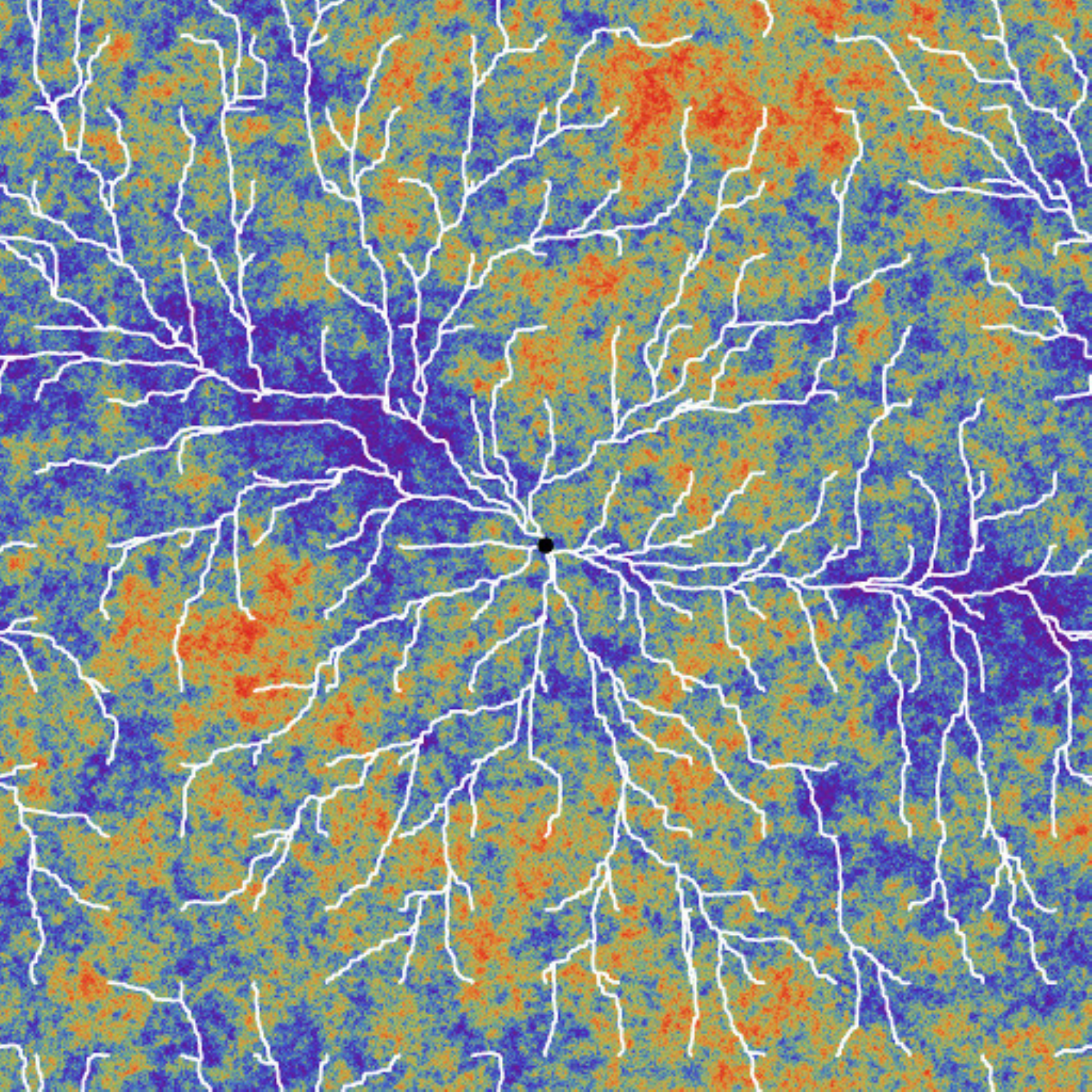}}\hfill
	\subfloat[$\xi=0.4$]{\includegraphics[width=.3\linewidth]{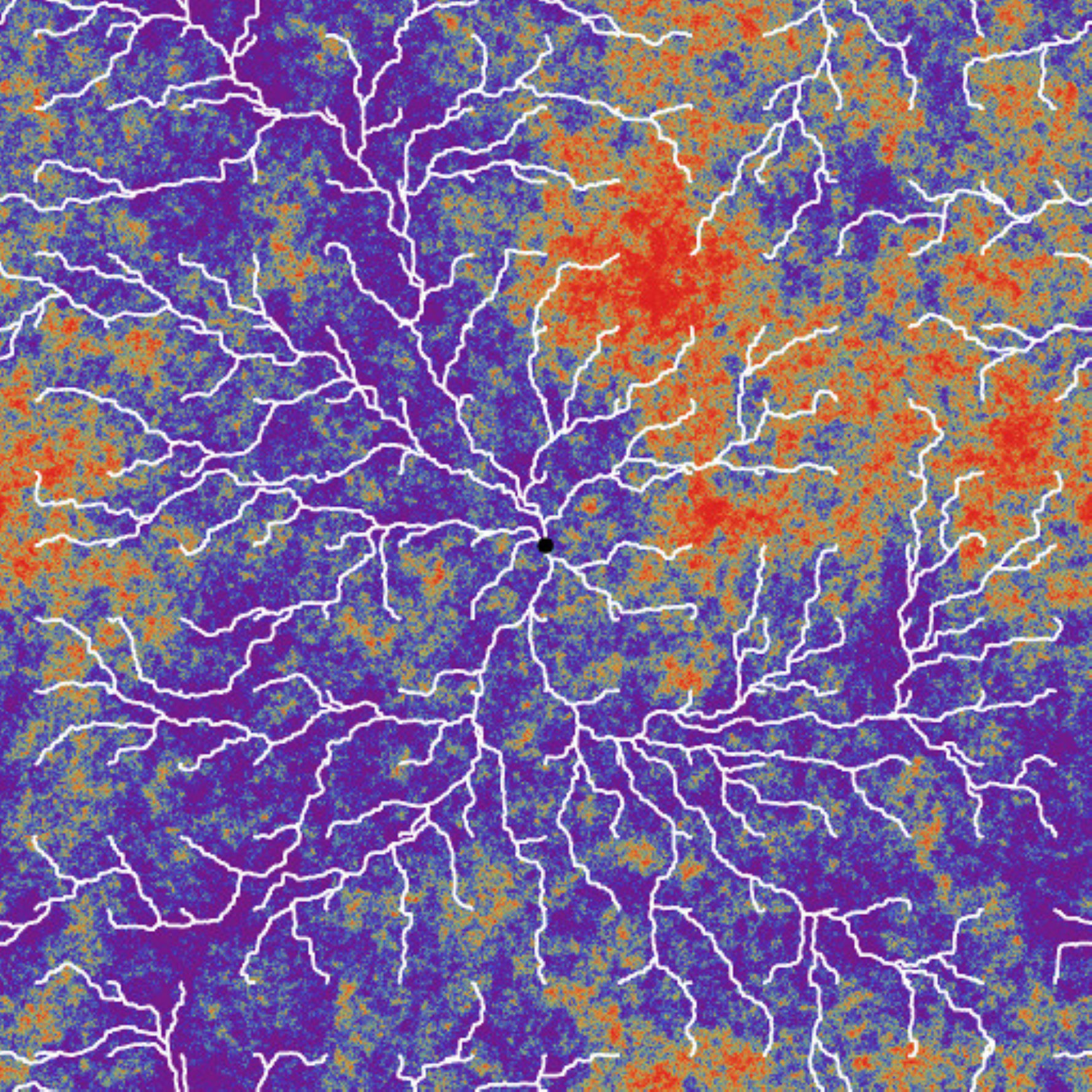}}
	
	\subfloat[$\xi=0.1$]{\includegraphics[width=.3\linewidth]{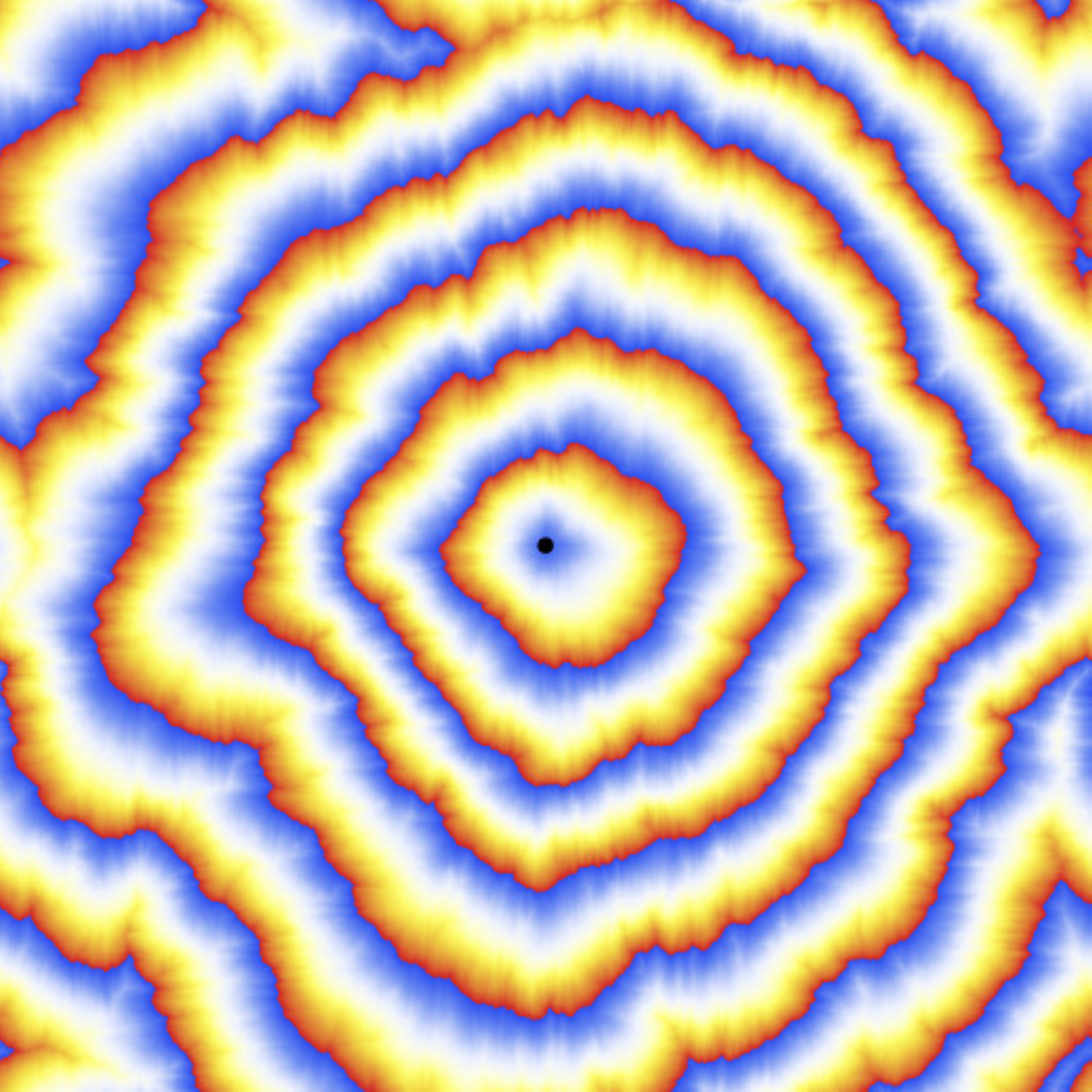}}\hfill
	\subfloat[$\xi=0.25$]{\includegraphics[width=.3\linewidth]{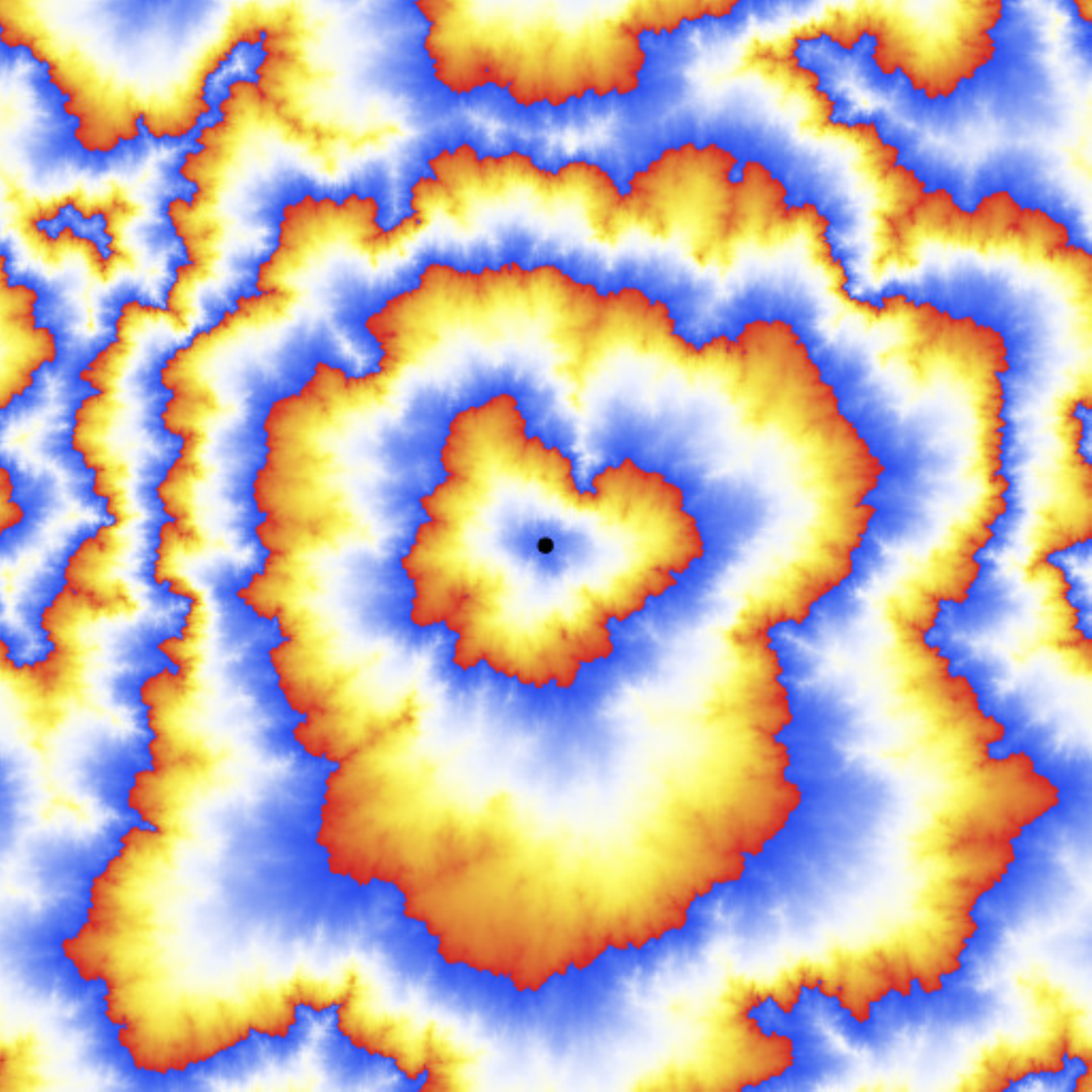}}\hfill
	\subfloat[$\xi=0.4$]{\includegraphics[width=.3\linewidth]{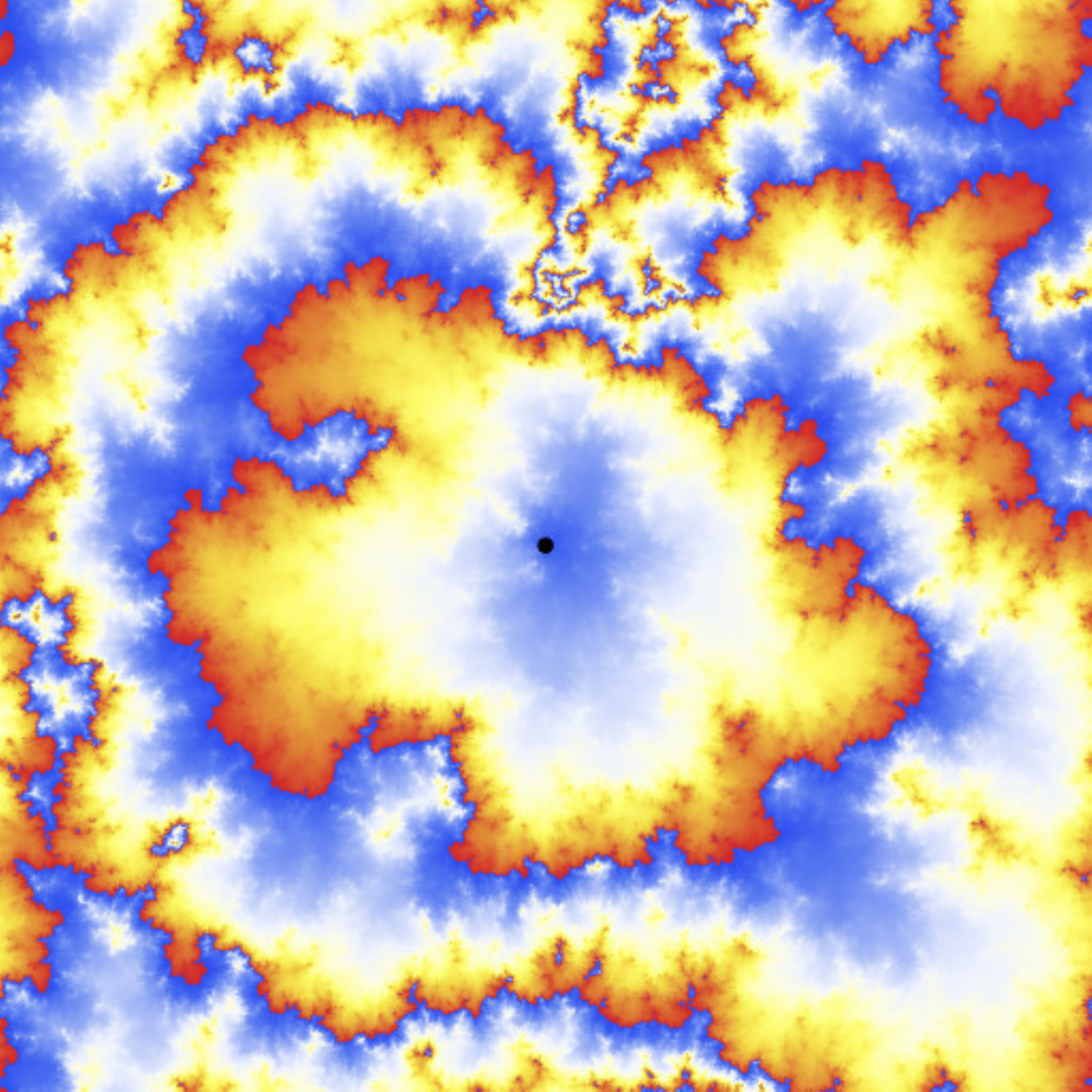}}
	\caption{Top: plot of $\xi \psi(x)$ on $\Lambda_{2048}$ together with the shortest DLFPP paths (white) from $15^2$ evenly spaced points to the center of the lattice. Bottom: the DLFPP distance $D_{\xi,w}(x,y)$ from points $x\in\Lambda_{2048}$ to the center $y$, colored from blue to red according to the fractional part of $\frac{10}{w}D_{\xi,w}(x,y)$. Note: all images use different samples of the discrete Gaussian free field. \label{fig:fieldplot}}
\end{figure}

The Liouville first passage percolation distance has a natural discrete counterpart \cite{Benjamini2010,Ding2018a,Ang2019} that is particularly convenient for numerical simulations.
Consider a $w\times w$ square lattice $\Lambda_w\coloneq(\Z/w\Z)^2$ with periodic boundary conditions.
The \emph{discrete Gaussian free field} $\psi:\Lambda_w \to \R$ on this lattice has probability distribution proportional to
\begin{equation}
\exp\Big[-\frac{1}{4\pi} \sum_{x,y\in\Lambda_w} \psi(x) \Delta_{xy}\psi(y) \Big]\delta\Big(\sum_{x\in\Lambda_w}\psi(x)\Big),\quad \Delta_{xy}=\begin{cases}
4 & x=y\\
-1 & \text{$x$ and $y$ adjacent}\\
0 & \text{otherwise,}
\end{cases}
\end{equation}
which is the natural discrete analogue of $e^{-S_{\mathrm{GFF}}[\phi]}$ in \eqref{eq:gffaction}. 
The normalization of the field is such that 
\begin{equation}
\langle\psi(0)^2\rangle \stackrel{w\to\infty}{\sim} \log w.
\end{equation}
The natural discretization of the first passage percolation distance is
\begin{equation}
D_{\xi,w}(x,y) = \inf_{\Gamma : x \to y} \sum_{i=1}^{m} \tfrac{1}{2}\left(e^{\xi \psi(\Gamma(i-1))}+e^{\xi \psi(\Gamma(i))}\right), 
\end{equation}
where the sum is over nearest-neighbour walks $\Gamma : \{0,1,\ldots m\} \to \Lambda_w$ of arbitrary length $m$ from $\Gamma(0) = x$ to $\Gamma(m) = y$. 
See Figure \ref{fig:fieldplot} for a few random samples.

In \cite[Theorem 1.4]{Ang2019} it is shown, in the slightly different setting of Dirichlet instead of periodic boundary conditions, that this distance approximates the continuum first passage percolation distance \eqref{eq:fpp} well.
In particular, it satisfies the analogous scaling relation \cite[Theorem 1.5]{Ang2019}
\begin{equation}\label{eq:dlfppscaling}
\frac{\log D_{\xi,w}([w\, x],[w\, y])}{\log w} \xrightarrow{w\to \infty} 1-\lambda(\xi)
\end{equation}
for $x,y\in[0,1)^2$ fixed, where $[w x]$ denotes the lattice point in $\Lambda_w$ closest to $w x$.

\section{Finite-size scaling of Liouville first passage percolation distance}\label{sec:fsslqg}
\begin{figure}[t]
	\centering
	\includegraphics[width=\linewidth]{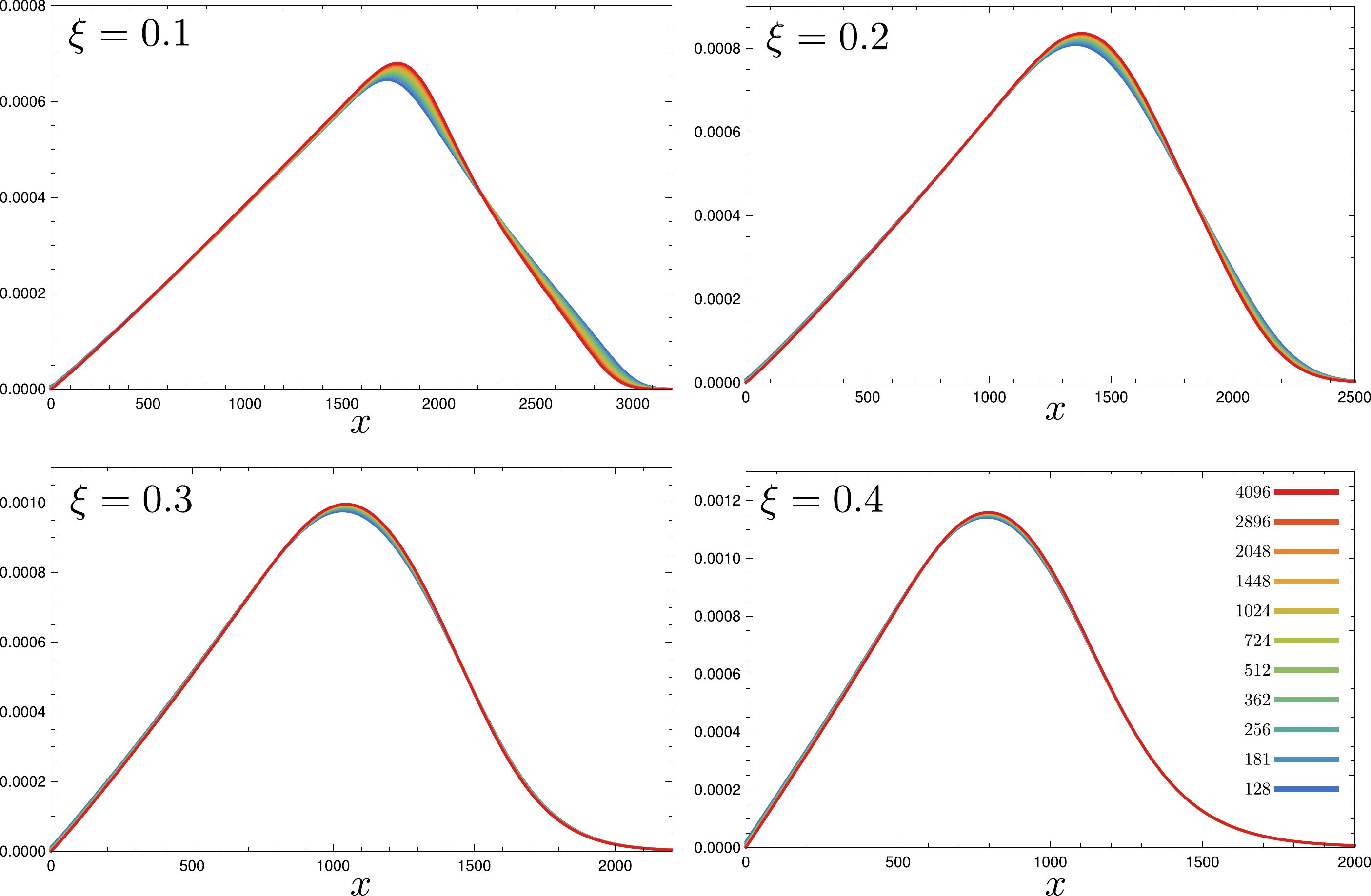}
	\caption{Finite-size scaling of the DLFPP distance for $\xi = 0.1, 0.2, 0.3, 0.4$. Shown are plots of $x\mapsto k_w^{-1} \rho_{\xi,w}(k_w^{-1}x)$ with $k_w$ determined by a best fit for $w = 2^7, \ldots, 2^{12}$. \label{fig:fpp_collapse_simple}}
\end{figure}
We make a similar assumption as we did in the case of the random planar maps, namely that the probability density $\rho_{\xi,w}$ of the distance $D_{\xi,w}(x,y)$ between two points $x$ and $y$ sampled uniformly from $\Lambda_w$ satisfies a pointwise scaling limit
\begin{equation}\label{eq:fppscalingansatz}
\lim_{w\to\infty} w^{1-\lambda} \rho_{\xi,w} (w^{1-\lambda} r) = \rho_\xi(r), \quad r>0.
\end{equation}
To estimate $\rho_{\xi,w}$ we have sampled (at least) $2$ million instances of the Gaussian free field for $w$ ranging from $2^6$ to $2^{12}$ and $\xi$ from $0.01$ to $0.4$.
For each field we pick an arbitrary starting point $x$ and determine the distances $D_{\xi,w}(x,y)$ to all points $y\in \Lambda_w$, which are then included in a histogram.

As in the case of the planar maps, we fit $x\mapsto k_w^{-1}\rho_{\xi,w}(k_w^{-1}x)$ to the reference distribution $\rho_{\xi,w_0}(x)$, where $w_0 = 2^{12}$ is the largest lattice size considered.
As before, only the data for which $\rho_{\xi,w}(r) \geq \nu \max_{r'}\rho_{\xi,w}(r')$ with $\nu = 0.2$ is used in the fit.
Figure~\ref{fig:fpp_collapse_simple} plots $k_w^{-1}\rho_{\xi,w}(k_w^{-1}x)$ for the fitted values of $k_w$ and various values of $\xi$.
The quality of the collapse is good for the larger values of $\xi$, and can be further improved by introducing a constant shift as was done before. 

\begin{figure}
	\centering
	\vspace{1.1cm}
	\includegraphics[width=\linewidth]{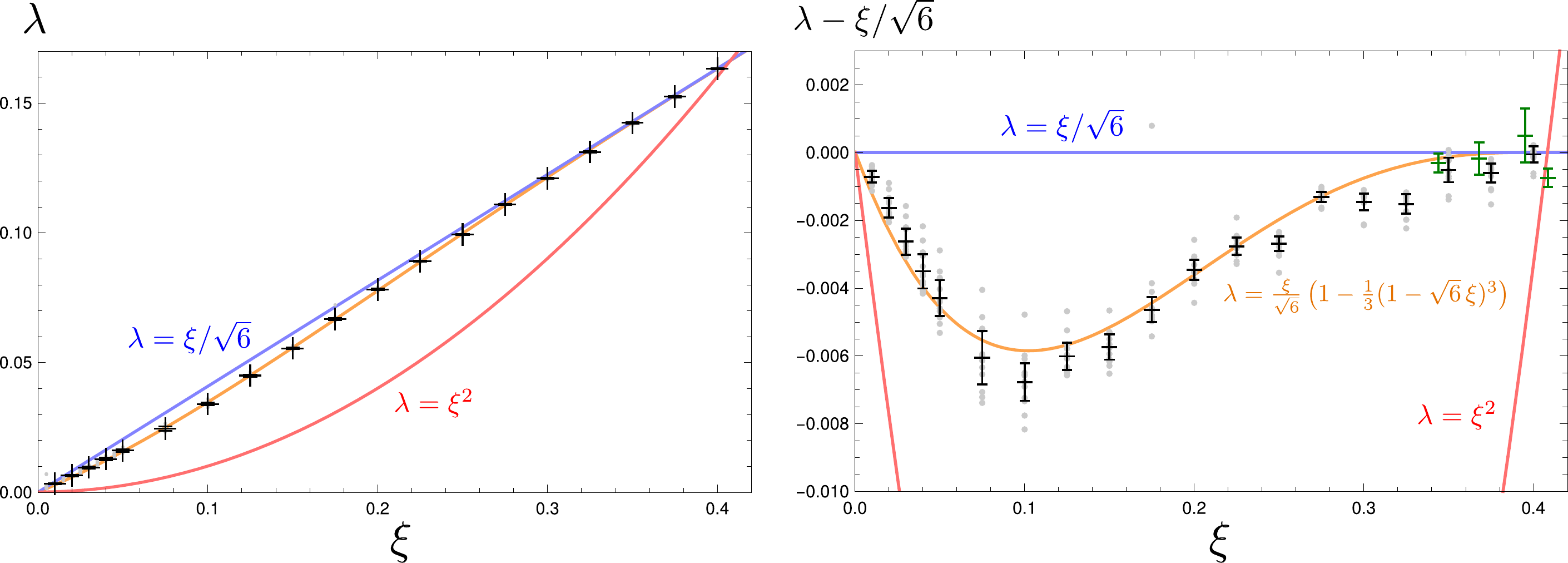}
	\caption{The plot on the left shows the estimates of $\lambda$ with systematic errors (as recorded in Table~\ref{tbl:lambda}) in comparison to $\lambda^{\mathrm{W}} = \xi^2$ (red), $\lambda^{\mathrm{DG}} = \xi/\sqrt{6}$ (blue) and $\lambda = \frac{\xi}{\sqrt{6}}\left(1-\frac13(1-\sqrt{6}\,\xi)^3\right)$ (orange). The right plot shows the same data normalized by $\xi/\sqrt{6}$, as well as the individual data points for the different parameters $\nu$ and $s$ (gray) and the planar map estimates of Table~\ref{tbl:pmdimensions} (green). \label{fig:lambda_estimate}}
\end{figure}

For small values of $\xi$, however, the approach towards the limiting distribution $\rho_\xi$ of \eqref{eq:fppscalingansatz} is markedly slower.
In this regime the fitted values of $k_w$ depend more sensitively on the fitting procedure used and one should attribute a larger systematic uncertainty to them.
To get a handle on this uncertainty we repeat the analysis with different choices of fitting parameters, namely $\nu = 0, 0.2, \ldots, 0.8$ for the range of data used and $s = 0, 2, 4$ for the constant shift (covering roughly the range of optimal shifts).
For each choice of these parameters the values $k_w$ are determined and fitted to the ansatz
\begin{equation}\label{eq:kwansatz}
k_w \approx \left(\frac{w}{w_0}\right)^{\lambda-1}\left(a + b \left(\frac{w}{w_0}\right)^{-\delta}\right)
\end{equation}
analogous to \eqref{eqn:kfitleadingorder}.
The collection of values of $\lambda$ obtained in this way is used to establish the systematic error on our estimate, which turns out to be significantly larger than the statistical error for all $\xi \lesssim 0.35$.

\begin{figure}[t]
	\centering
	\includegraphics[width=\linewidth]{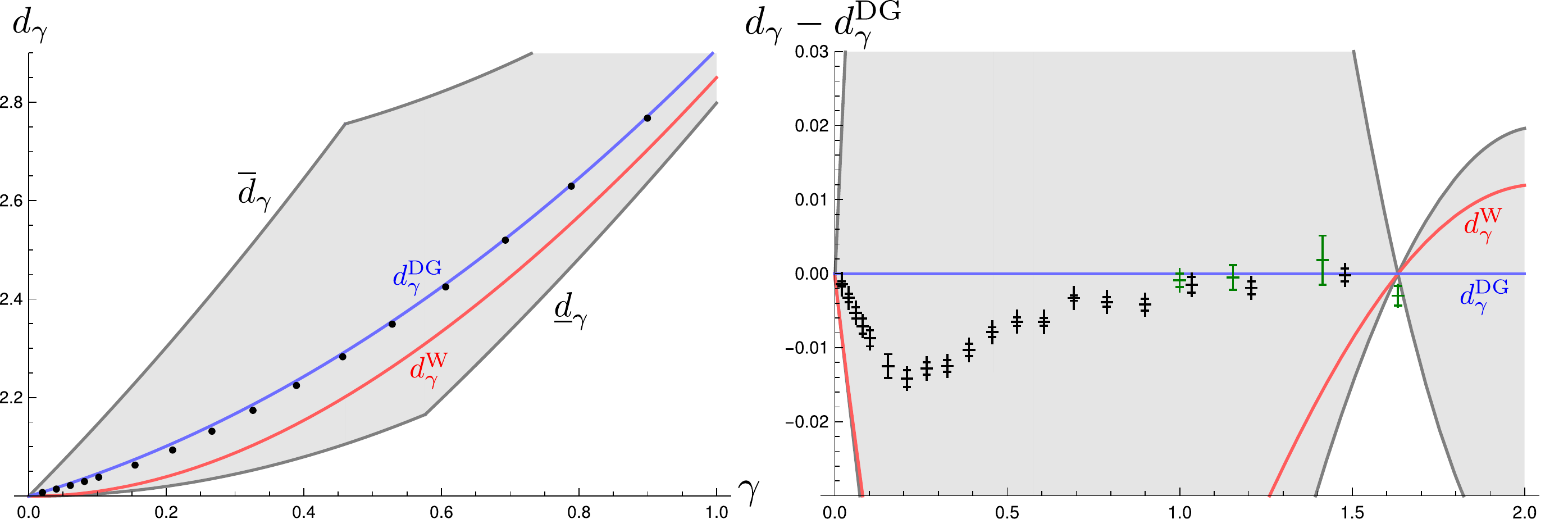}
	\caption{Left: The DLFPP estimates of $d_\gamma$ (error bars too small to display). Right: the estimates of $d_\gamma$ normalized by $d_\gamma^{\mathrm{DG}}$ with error bars as obtained from DLFPP (black) and random planar maps (green).  \label{fig:lfpp_dimensions}}
\end{figure}

The results are gathered in Table~\ref{tbl:lambda}, which also includes the corresponding central charge $c= 25- 6(1-\lambda)^2/\xi^2$ and estimates for $\gamma$ and $d_\gamma$ calculated using \eqref{eq:xi} and \eqref{eq:lambda}.
We only record the error in $d_\gamma$ explicitly, because it is most significant when comparing to the formulas $d_{\gamma}^{\mathrm{DG}}$ and $d_\gamma^{\mathrm{W}}$.
Figure~\ref{fig:lambda_estimate} shows a plot of the estimated values of $\lambda$, while Figure~\ref{fig:lfpp_dimensions} displays the corresponding estimates for the Hausdorff dimension $d_\gamma$.

\begin{table}[!t]
	\caption{\label{tbl:lambda} Estimates for $\lambda$ with systematic errors obtained by fitting to \eqref{eq:kwansatz}.}
	\begin{indented}
		\item[]
	\begin{tabular}{ccllc}\hline
	$\xi$ & $\lambda$  & $\quad c$ & $\quad\gamma$ & $d_\gamma$  \\\hline
	0.010 & $\quad 0.0033 \pm 0.0002 \quad$ & $-59.6\mathrm{k}\quad$ & $0.0201\quad$ & $2.0070 \pm 0.0003$ \\
	0.020 & $0.0065 \pm 0.0003$ & $-14.8\mathrm{k}$ & 0.0402 & $2.0140 \pm 0.0006$ \\
	0.030 & $0.0096 \pm 0.0004$ & $-6.5\mathrm{k}$ & 0.0606 &  $2.0213 \pm 0.0008$ \\
	0.040 & $0.0128 \pm 0.0005$ & $-3.6\mathrm{k}$ & 0.0812 & $2.0293 \pm 0.0010$ \\
	0.050 & $0.0161 \pm 0.0005$ & $-2.3\mathrm{k}$ & 0.1019 & $2.0380 \pm 0.0011$ \\
	0.075 & $0.0246 \pm 0.0008$ & $-990.$ & 0.1547 & $2.0626 \pm 0.0017$ \\
	0.100 & $0.0341 \pm 0.0006$ & $-534.$ & 0.2093 & $2.0932 \pm 0.0012$ \\
	0.125 & $0.0450 \pm 0.0004$ & $-325.$ & 0.2664 & $2.1315 \pm 0.0009$ \\
	0.150 & $0.0555 \pm 0.0004$ & $-213.$ & 0.3261 & $2.1738 \pm 0.0009$ \\
	0.175 & $0.0668 \pm 0.0004$ & $-145.$ & 0.3893 & $2.2244 \pm 0.0010$ \\
	0.200 & $0.0782 \pm 0.0003$ & $-102.$ & 0.4565 & $2.2827 \pm 0.0008$ \\
	0.225 & $0.0891 \pm 0.0003$ & $-73.3$ & 0.5285 & $2.3489 \pm 0.0007$ \\
	0.250 & $0.0994 \pm 0.0002$ & $-52.9$ & 0.6062 & $2.4247 \pm 0.0007$ \\
	0.275 & $0.1110 \pm 0.0001$ & $-37.7$ & 0.6929 & $2.5196 \pm 0.0005$ \\
	0.300 & $0.1210 \pm 0.0003$ & $-26.5$ & 0.7888 & $2.6293 \pm 0.0010$ \\
	0.325 & $0.1312 \pm 0.0003$ & $-17.9$ & 0.8994 & $2.7675 \pm 0.0014$ \\
	0.350 & $0.1424 \pm 0.0004$ & $-11.0$ & 1.0346 & $2.9561 \pm 0.0022$ \\
	0.375 & $0.1525 \pm 0.0003$ & $-5.65$ & 1.2076 & $3.2201 \pm 0.0023$ \\
	0.400 & $0.1632 \pm 0.0002$ & $-1.26$ & 1.4787 & $3.6968 \pm 0.0035$ \\\hline
	\end{tabular}
	\end{indented}
\end{table}

\section{Discussion}
The DLFPP estimates of $\lambda$ and $d_\gamma$ for $\xi = 0.35,\, 0.375,\, 0.4$ are in very good agreement with Ding \& Gwynne's prediction $\lambda^{\mathrm{DG}} = \xi / \sqrt{6}$ and consistent with the random planar map results (see the green data points in Figure~\ref{fig:lambda_estimate} and Figure~\ref{fig:lfpp_dimensions}).  
For $\xi < 0.35$ the measurements are still much closer to $\lambda^{\mathrm{DG}}$ than to Watabiki's prediction $\lambda^{\mathrm{W}} = \xi^2$, but a significant negative deviation is visible, that is most pronounced at around $\xi = 0.1$ ($\gamma\approx 0.2$, $c \approx -2.3k$). 
The data is better described by the (completely ad hoc) formula $\lambda = \frac{\xi}{\sqrt{6}}\left(1-\frac13(1-\sqrt{6}\,\xi)^3\right)$.
However, we are hesitant to rule out $\lambda^{\mathrm{DG}} = \xi / \sqrt{6}$ on the basis of the current data.
The reason for this is that the quality of the finite-size scaling for smaller values of $\xi$ is not as good as one may have hoped, indicating that (much) larger lattice sizes might be necessary to observe accurate scaling.
An explanation could be that for small $\xi$ and currently used lattice sizes, the DLFPP geodesics are too close to geodesics of the Euclidean lattice and therefore do not sufficiently display their fractal structure.
Indeed, in Figure \ref{fig:fieldplot} many of the DLFPP geodesics for $\xi=0.1$ are seen to contain fairly long straight segments, a phenomenon that becomes more pronounced for even smaller $\xi$. 
Since the DLFPP length of straight segments scales with an exponent that is different but for small $\xi$ quite close to $1-\lambda$, one may need to increase the lattice size $w$ considerably to make its subleading contribution small enough.

\section*{Source code and simulation data}

The source code and simulation data, on which Section \ref{sec:fssplanarmaps} and Section \ref{sec:fsslqg} are based, are freely available online at \cite{Barkley2019}.

\section*{References}

\providecommand{\newblock}{}

\end{document}